\documentstyle[amssymb,eqsecnum,aps,epsfig]{revtex}

\newfont{\DamirFont}{arxi}
\newfont{\blackboard}{msbm10}%%%%%%%%%% scaled\magstep2}
\newcommand{\Z}{\mbox{\blackboard\symbol{"5A}}}
\def\Res{\mathop{\rm Res}}
\def\equals{\mathop{=}}
\newcommand{\io}{[\hspace{-1pt}[}
\newcommand{\ic}{]\hspace{-1pt}]}
\newcommand{\fo}{\{\!\mid\!}
\newcommand{\fc}{\!\mid\!\}}
\newcommand{\bGamma}{\mbox{\boldmath $\Gamma$}}

\newcommand{\sgn}{{\rm sgn}}
\newcommand{\Det}{{\rm Det}}
\newcommand{\BS}{{\rm BS}}
\newcommand{\Fi}{{\cal F}}
\newcommand{\E}{{\cal E}}
\renewcommand{\Re}{{\rm Re\,}}
\newcommand{\REG}{{\rm REG}}
\newcommand{\IRREG}{{\rm IRREG}}

\newcommand{\ren}{{\rm ren}}
\newcommand{\eff}{{\rm eff}}
\newcommand{\tr}{{\rm tr}}

\renewcommand{\tanh}{{\rm tanh}}
\renewcommand{\arctan}{{\rm arctan}}

\renewcommand{\cosh}{{\rm cosh}}

\newcommand{\Lp}{L_{(+)}}
\newcommand{\Lm}{L_{(-)}}
\newcommand{\tE}{\tilde{E}}
\newcommand{\tLp}{\tilde{L}_{(+)}}
\newcommand{\tLm}{\tilde{L}_{(-)}}
\newcommand{\Rp}{R_{(+)}^{(+)}}
\newcommand{\Rm}{R_{(-)}^{(-)}}
\newcommand{\Rpm}{R_{(-)}^{(+)}}
\newcommand{\Rmp}{R_{(+)}^{(-)}}
\newcommand{\tRp}{\tilde{R}_{(+)}^{(+)}}
\newcommand{\tRm}{\tilde{R}_{(-)}^{(-)}}
\newcommand{\tRpm}{\tilde{R}_{(-)}^{(+)}}
\newcommand{\tRmp}{\tilde{R}_{(+)}^{(-)}}

\def\be{\begin{equation}}
\def\ee{\end{equation}}
\def\bea{\begin{eqnarray}}
\def\eea{\end{eqnarray}}
\def\ba{\begin{array}}
\def\ea{\end{array}}
\def\V{{\bf V}}
\def\J{{\bf J}}
\def\j{{\bf j}}

\def\vac{{\rm vac}}
\def\x{{\bf x}}
\def\F{\Phi^{(0)}}
\def\M{{\cal M}}
\def\N{{\cal N}}
\def\S{{\cal S}}
\def\L{{\cal L}}
\def\A{{\cal A}}
\def\C{{\cal C}}

\def\i{{\int}}

\newcommand{\bpar}{\mbox{\boldmath $\partial$}}
\newcommand{\ab}{\mbox{\boldmath $\alpha$}}
\newcommand{\gb}{\mbox{\boldmath $\gamma$}}

\def\si{\mathop{\displaystyle\sum\mkern-25mu\int\,}}

\newcommand{\ds}{\displaystyle}
\newcommand{\sss}{\scriptscriptstyle}

\begin{document}
\title{\bf Self-adjointness of the two-dimensional massless
Dirac Hamiltonian and vacuum polarization effects in 
the background of a singular magnetic vortex}

\author{{\large\bf Yurii A. Sitenko\thanks{Electronic address: yusitenko@bitp.kiev.ua}}\\
Bogolyubov Institute for Theoretical Physics, National Academy of
Sciences,\\14-b Metrologichna str., 03143 Kyiv, Ukraine
\thanks{Permanent address}\\
and\\
Institute for Theoretical Physics, Berne University,\\
Sidlerstrasse 5, 3012 Berne, Switzerland\\}

\maketitle
\begin{abstract}
A massless spinor field is quantized in the background of a singular 
static magnetic vortex in 2+1-dimensional space-time. The method of 
self-adjoint extensions is employed to define the most general set 
of physically acceptable boundary conditions at the location of the 
vortex. Under these conditions, all effects of polarization of the 
massless fermionic vacuum in the vortex background are determined. 
Absence of anomaly is demonstrated, and patterns of both parity and 
chiral symmetry breaking are discussed.

\smallskip
\noindent PACS number(s): 03.65.Bz, 12.20.Ds, 11.30.Er, 11.30.Qc 
\end{abstract}

\section{Introduction}
Singular (or contact or zero-range) interaction potentials were
introduced in  quantum mechanics more than sixty years ago \cite{Bet,Tho,Fer}.
Since that time the attitude of physicists and mathematicians to this subject
was varying, starting from "it is impossible", then to "it is evident", and
finally arriving at "it is interesting" (for a review see monograph 
\cite{Alb}). A mathematically consistent and rigorous treatment of the subject 
was developed \cite{Ber}, basing on the notion of self-adjoint extension
of a Hermitian (symmetric) operator.

 Singular interaction is involved in quantum field theory when, for example,
a spinor field is quantized in the background of a pointlike magnetic monopole
in threedimensional space or a pointlike magnetic vortex in twodimensional
space. In these cases the Dirac Hamiltonian, in contrast to the Schrodinger one,
is free from an explicit $\delta$-function singularity;
nonetheless the problem of self-adjoint extension of both Dirac and Schrodinger
operators arises, albeit for different reasons (see, for example \cite{Gacki}).
A distinguishing feature is that a
solution to the Dirac equation, unlike that to the Schrodinger one, cannot 
obey a condition of regularity at the singularity point. It is
necessary then to define a boundary condition at this point, and the
least restrictive, but still physically acceptable, condition is such that
guarantees self-adjointness of the Dirac Hamiltonian. Thus, effects of
polarization of the fermionic vacuum in a singular background
(such as a pointlike monopole or a pointlike vortex) appear to
depend on the choice of the boundary condition at the singularity
point, and a set of permissible boundary conditions is labelled, most
generally, by the values of self-adjoint extension parameters.
In contrast to the Schrodinger case, the extension in the Dirac case does not
reflect additional types of interaction but represents complementary
information that must be specified when describing the physical attributes
of the already posited singular background configuration.

As a consequence, the fermionic vacuum under the influence of a
singular background can acquire rather unusual properties: leakage of
quantum numbers from the singularity point  occurs. While in the case
of a monopole there
is leakage of charge to the vacuum, which results in the monopole
becoming the dyon violating the Dirac quantization condition and CP
symmetry \cite{Gol,Cal,Wit,Gro,Yam}, in the case
of a vortex (the Ehrenberg-Siday-Aharonov-Bohm potential \cite{Ehre,Aha})
the situation is much
more complicated, since there is leakage of both charge and other
quantum numbers to the vacuum.
Apparently, this is due to a nontrivial topology of the base space in the
latter case: $\pi_1=0$ in the case of a space with a deleted point, and
$\pi_1=\Z$ in the case of a space with
a deleted line (or a plane with a deleted point); here $\pi_1$ is the
first homotopy group and $\Z$ is a set of integer numbers.
For a particular choice of the boundary
condition at the location of a singular vortex it has been shown that charge
\cite{Sit88,Sit90}, current \cite{Gor} and angular momentum
\cite{SitR96} are induced in the vacuum. The induced vacuum quantum
numbers under general boundary conditions which are  compatible with
self-adjointness have been considered in Refs.\cite{Sit96,Sit97,SitR97,Sit99}.

Thus far the effects of polarization of the massive fermionic vacuum
have been studied.
In a 2+1-dimensional space-time the mass term for a spinor field
in an irreducible representation of the Clifford algebra
violates both types of parity  --  under space and time reflections.
One can also consider a third type of parity, which is similar
to the axial symmetry in evendimensional space-times (see Refs.
\cite{Nie,Alva1,Redl,Alva2}) and which is violated by the mass
term as well.

When quantization of a massless spinor field is
considered, all above symmetries are formally present. Our
concern is then, in the first place, in the following: whether the vacuum 
polarization effects in a singular background respect these
formal symmetries?
It will be shown that, although the parity anomaly  is absent, the
parity breaking condensate emerges  in the vacuum.
If the number of quantized spinor fields is doubled to form
a reducible representation composed of two inequivalent irreducible ones,
then this condensate becomes parity conserving but chiral symmetry
breaking. Also all
other characteristics  of the massless fermionic vacuum in a singular
background are determined.

A singular background in $2+1$ dimensional space-time is taken in the
form of a pointlike static magnetic vortex (the Ehrenberg-Siday-Aharonov-Bohm
configuration)
$$V^1(\x)=-\Phi^{(0)}{x^2\over(x^1)^2+(x^2)^2}, \quad V^2(\x)=\F
{x^1\over(x^1)^2+(x^2)^2}, \eqno(1.1)$$
$$\bpar\times\V(\x)=2\pi\F\delta(\x),
\eqno(1.2) $$
where $\F$ is the vortex flux in $2\pi$ units, i.e.
in the London ($2\pi \hbar c e^{-1}$) units, since we use
conventional units $\hbar=c=1$ and coupling constant $e$ is included into
vector potential $\V(\x)$.
The wave function on the plane ($x^1$, $x^2$) with punctured singular
point $x^1=x^2=0$
obeys the most general condition (see \cite{Sit96} for more details)
$$<r,\varphi+2\pi|=e^{i2\pi\Upsilon}<r,\varphi| \,  , \eqno(1.3)$$
where $r=\sqrt{(x^1)^2+(x^2)^2}$ and $\varphi=\arctan(x^2/x^1)$ are the
polar coordinates, and $\Upsilon$ is a continuous real parameter which
is varied in the range $0\leq\Upsilon<1$. It can be shown (see, for
example, \cite{SitR96,Sit96}) that $\Upsilon$ as well as $\F$ is
changed under singular gauge transformations, whereas difference
$\F-\Upsilon$ remains invariant. Thus, physically sensible quantities
are to depend on the gauge invariant combination $\F-\Upsilon$ which
will be for brevity denoted as the reduced vortex flux in the following.

In the next section, the most general set of physically acceptable boundary
conditions at the singularity point $\x=0$ is defined. In Section III, the 
vacuum fermion number is determined. Section IV is central from the technical 
point of view, since we show here, how the problem of both ultraviolet and 
infrared divergences in vacuum characteristics is solved with the help of zeta 
function regularization. This allows us to get immediately
in Section V the vacuum energy density; also here the effective action and the
effective potential are considered. The vacuum current is determined in Section
VI, and the vacuum condensate is determined in Section VII. We demonstrate
in Section VIII that the parity anomaly is absent in the background of a
singular magnetic vortex. Section IX is devoted to the determination
of the last vacuum characteristics -- angular momentum. We summarize
results and discuss their consequences in Section X. Some crucial points 
of the derivation of results are outlined in Appendices A-D.

\section{Quantization of a Spinor Field and the Boundary Condition at the
Location of a Vortex}

The operator of the second-quantized spinor field is presented in the
form
$$
\Psi(\x,t)=\si_{E_\lambda>0}\, e^{-iE_\lambda
t}<\x|\lambda>a_\lambda+\si_{E_\lambda<0}\, e^{-iE_\lambda
t}<\x|\lambda>b_\lambda^+, \eqno(2.1)
$$
where $a_\lambda^+$ and $a_\lambda$ ($b_\lambda^+$ and $b_\lambda$) are
the spinor particle (antiparticle) creation and annihilation operators
satisfying anticommutation relations
$$
[a_\lambda,a_{\lambda'}^+]_+=[b_\lambda,b_{\lambda'}^+]_+=<\lambda|\lambda'>,
\eqno(2.2)
$$
and $<\x|\lambda>$ is the solution to the stationary Dirac equation
$$ H<\x|\lambda>=E_\lambda<\x|\lambda>, \eqno(2.3)$$
$H$ is the Dirac Hamiltonian, $\lambda$ is the set of parameters
(quantum numbers) specifying a state, $E_\lambda$ is the energy of a 
state; symbol $\si$ means the summation over discrete and the
integration (with a certain measure) over continuous values of
$\lambda$. The ground state $|\vac>$ is defined conventionally by
equality
$$ a_\lambda|\vac>=b_\lambda|\vac>=0. \eqno(2.4) $$
In the case of quantization of a massless spinor field in the
background of static vector field $\V(\x)$, the Dirac Hamiltonian
takes the form
$$ H=-i\ab[\bpar-i\V(\x)], \eqno(2.5)$$
where
$$\ab=\gamma^0\gb, \qquad \beta=\gamma^0, \eqno(2.6)$$
$\gamma^0$ and $\gb$ are the Dirac $\gamma$ matrices. In the 
2+1-dimensional space-time $(\x,t)=(x^1,x^2,t)$ the Clifford algebra
has two inequivalent irreducible representations which can be differed
in the following way:
$$i\gamma^0\gamma^1\gamma^2=s, \qquad s=\pm1.
\eqno(2.7)$$
Choosing the $\gamma^0$ matrix in the diagonal form
$$\gamma^0=\sigma_3 , \eqno(2.8)$$
one gets
$$\gamma^1=e^{{i\over2}\sigma_3\chi_s}i\sigma_1e^{-{i\over2}\sigma_3\chi_s},
\quad \gamma^2=
e^{{i\over2}\sigma_3\chi_s}is\sigma_2e^{-{i\over2}\sigma_3\chi_s},
\eqno(2.9)$$
where $\sigma_1,\sigma_2$ and $\sigma_3$ are the Pauli matrices,
and $\chi_1$ and
$\chi_{-1}$ are the parameters that are varied in the interval
$0\leq\chi_s<2\pi$ to go over to the equivalent representations.

A solution to the Dirac equation (2.3) with Hamiltonian (2.5) in 
background (1.1), that obeys condition (1.3), can be
presented as
$$<\x|E,n>=\left(\ba{l}
f_n(r,E)e^{i(n+\Upsilon)\varphi}\\[0.2cm]
g_n(r,E)e^{i(n+\Upsilon+s)\varphi}\\ \ea \right), \quad n\in\Z,
\eqno(2.10)$$
where the column of radial functions
$\left(\ba{l}
f_n\\
g_n\\ \ea \right)$
satisfies the equation
$$h_n\left(\ba{l}
f_n\\[0.2cm]
g_n\\ \ea \right)=
E\left(\ba{l}
f_n\\[0.2cm]
g_n\\ \ea \right),
\eqno(2.11)$$
and
%$$
%h_n=
%\left(\ba{l}
%0\quad e^{i\chi_s}[\partial_r+s(n-\F+\Upsilon+s)r^{-1}] \\[0.2cm]
%e^{-i\chi_s}[-\partial_r+s(n-\F+\Upsilon)r^{-1}] \quad 0\\ \ea \right)
%\eqno(2.12)$$
$$
h_n=
\left(
\ba{l}
\quad \quad \quad \quad \quad \quad \quad 0\\[0.2cm]
e^{-i\chi_s}[-\partial_r+s(n-\F+\Upsilon)r^{-1}] \\
\ea
\ba{l}
e^{i\chi_s}[\partial_r+s(n-\F+\Upsilon+s)r^{-1}]\\[0.2cm]
\quad \quad \quad \quad \quad \quad \quad 0 \\
\ea
\right)
\eqno(2.12)$$
is the partial Dirac Hamiltonian.
When reduced vortex flux $\F-\Upsilon$ is integer, the requirement of
square integrability for wave function (2.10) at $r\rightarrow 0$ 
provides its regularity, rendering the partial Dirac
Hamiltonian $h_n$ for every value of $n$ to be essentially self-adjoint.
When $\F-\Upsilon$ is fractional, the same is valid only for $n\neq
n_0$, where
$$n_0=\io\F-\Upsilon\ic+{1\over2}-{1\over2}s, \eqno(2.13)$$
$\io u\ic$ is the integer part of a quantity $u$ (i.e., the greatest
integer that is less than or equal to $u$). For $n=n_0$, each of the
two linearly independent solutions to Eq.(2.11) meets the
requirement of square integrability at $r\rightarrow 0$. Any particular 
solution in this
case is characterized by at least one (at most both) of the radial
functions being divergent as $r^{-p}$ ($p<1$) at $r\rightarrow 0$. If
one of the two linearly independent solutions is chosen to have a
regular upper and an irregular lower component, then the other one has
a regular lower and an irregular upper component. Therefore,
in contrast to operator $h_n$ ($n\neq n_0$), operator
$h_{n_0}$ is not essentially self-adjoint
\footnote{A corollary of the
theorem proven in Ref.\cite{Wei} states that, for the partial Dirac
Hamiltonian to be essentially self-adjoint, it is necessary and
sufficient that a non-square-integrable (at $r\rightarrow 0$) solution exist.}.
The Weyl - von Neumann theory of self-adjoint operators (see, e.g.,
Refs.\cite{Alb,Akhie}) has to be employed in order to consider the
possibility of a self-adjoint extension in the case of $n=n_0$. It is shown
in Appendix A that the self-adjoint extension exists indeed and is
parametrized by one continuous real variable denoted in the following
by $\Theta$. Thus operator $h_{n_0}$
is defined on the domain of functions obeying the condition

$$\cos\bigl(s{\Theta\over2}+{\pi\over4}\bigr)\lim_{r\rightarrow 0}(\mu
r)^Ff_{n_0}=-e^{i\chi_s}\sin\bigl(s{\Theta\over2}+{\pi\over4}\bigr)
\lim_{r\rightarrow 0}(\mu r)^{1-F}g_{n_0}, \eqno(2.14)$$
where $\mu>0$ is the parameter of the dimension of inverse length and
$$
F=s\fo\F-\Upsilon\fc+{1\over2}-{1\over2}s,
\eqno(2.15)$$
$\fo u\fc=u-\io u\ic$ is the fractional part of a quantity $u$,
$0\leq\fo u\fc<1$; note here that Eq.(2.14) implies that $0<F<1$,
since in the case of $F={1\over2}-{1\over2}s$ both $f_{n_0}$ and
$g_{n_0}$ obey the condition of regularity at $r\rightarrow 0$. Note 
also that Eq.(2.14) is periodic in $\Theta$ with period $2\pi$; therefore,
without a loss of generality, all permissible values of $\Theta$ will be
restricted in the following to range $-\pi\leq\Theta\leq\pi$.

All solutions to the massless Dirac equation in the background of a singular
magnetic vortex correspond to the continuous spectrum and, therefore,
obey the orthonormality condition
$$
\int d^2x<E,n|\x><\x|E',n'>={\delta(E-E')\over\sqrt{|EE'|}}\delta_{nn'}.
\eqno(2.16)$$
In the case of $0<F<1$ one can get the following expressions
corresponding to the regular solutions with $sn>sn_0$:
$$
\left(\ba{c} f_n\\ g_n \\ \ea \right) ={1\over2\sqrt{\pi}}
\left(\ba{c}
J_{l-F}(kr)e^{i\chi_s}\\[0.2cm]
\sgn(E)J_{l+1-F}(kr)\\ \ea \right), \qquad l=s(n-n_0), \eqno(2.17)$$
the regular solutions with $sn<sn_0$:
$$
\left(\ba{c} f_n\\ g_n \\ \ea \right)
={1\over2\sqrt{\pi}}
\left(\ba{c}
J_{l'+F}(kr)e^{i\chi_s}\\[0.2cm]
-\sgn(E)J_{l'-1+F}(kr) \ea \right), \qquad l'=s(n_0-n), \eqno(2.18)$$
and the irregular solution:
$$
\left(\ba{c} f_{n_0}\\ g_{n_0} \\ \ea \right)
={1\over 2\sqrt{\pi[1+\sin(2\nu_E)\cos(F\pi)]}} \left(\ba{c}
[\sin(\nu_E)J_{-F}(kr)+\cos(\nu_E)J_F(kr)]e^{i\chi_s}\\[0.2cm]
\sgn(E)[\sin(\nu_E)J_{1-F}(kr)-\cos(\nu_E)J_{-1+F}(kr)]\\ \ea \right);
\eqno(2.19)$$
here $k=|E|$, $J_{\rho}(u)$ is the Bessel function of order
$\rho$ and
$$
\sgn(u)=\left\{\ba{cc}
1,& u>0\\
-1,& u<0\\ \ea \right\}.
$$
Substituting the asymptotic form of Eq.(2.19) at $r\rightarrow 0$
into Eq.(2.14), one arrives at the relation between parameters
$\nu_E$ and $\Theta$:
$$
\tan(\nu_E)=\sgn(E)\biggl({k\over2\mu}\biggr)^{2F-1}\,
{\Gamma(1-F)\over\Gamma(F)}\tan\bigl(s{\Theta\over2}+{\pi\over4}\bigr),
\eqno(2.20)$$
where $\Gamma(u)$ is the Euler gamma function.

Using the explicit form of solutions (2.17) -- (2.19), all vacuum
polarization effects can be determined.

\section{Fermion Number}

In the second-quantized theory in 2+1-dimensional space-time the
operator of the fermion number is given by the expression
$$
\hat{\N}=\int d^2x\,{1\over2}[\Psi^+(\x,t),\Psi(\x,t)]_-=\si
[a_\lambda^+a_\lambda-b_\lambda^+b_\lambda-{1\over2} \sgn(E_\lambda)],
\eqno(3.1)$$
and, consequently, its vacuum expectation value takes the form
$$
\N\equiv<\vac|\hat{\N}|\vac>=-{1\over2}\si
\sgn(E_\lambda)=-{1\over2}\int d^2x\,\tr<\x|\,\sgn(H)|\x>. \eqno(3.2)$$
From general arguments, one could expect that the last quantity
vanishes due to cancellation between the contributions of positive and
negative energy solutions to the Dirac equation (2.3). Namely this
happens in a lot of cases. That is why every case of a nonvanishing value
of $\N$ deserves a special attention.

Considering the case of the background in the form of a singular
magnetic vortex (1.1) -- (1.2), one can notice that the contribution
of regular solutions (2.17) and (2.18) is cancelled upon summation
over the sign of energy, whereas irregular solution (2.19) yields a
nonvanishing contribution to $\N$ (3.2). Defining the vacuum fermion
number density
$$ \N_{\x}=-{1\over2}\tr<\x|\,\sgn(H)|\x>, \eqno(3.3)$$
we get
$$
\N_\x=-{1\over8\pi}\int\limits_0^\infty dkk\biggl\{
A\biggl({k\over\mu}\biggr)^{2F-1}
\left[\Lp+\Lm\right]\left[J_{-F}^2(kr)
+J_{1-F}^2(kr)\right]+$$
$$+2\left[\Lp-\Lm\right]\left[J_{-F}(kr)J_F(kr)-
J_{1-F}(kr)J_{-1+F}(kr)\right]+$$
$$+A^{-1}\biggl({k\over\mu}\biggr)^{1-2F}\left[\Lp+\Lm\right]
\left[J_F^2(kr)+J_{-1+F}^2(kr)\right]\biggr\}, \eqno(3.4)$$
where
$$A=2^{1-2F}{\Gamma(1-F)\over\Gamma(F)}\tan\left(s{\Theta\over2}+
{\pi\over4}\right) \eqno(3.5)$$
and
$$L_{(\pm)}=2^{-1}\bigl\{\cos(F\pi)\pm\cosh\bigl[(2F-1)\ln({k\over\mu})+
\ln A\bigr]\bigr\}^{-1}. \eqno(3.6)$$
We show in Appendix B, how the integral in Eq.(3.4) is transformed resulting
in the expression
$$\N_\x=-{\sin(F\pi)\over2\pi^3r^2}\int\limits_0^\infty dw\,
w {K_F^2(w)-K_{1-F}^2(w)\over \cosh[(2F-1)\ln({w\over\mu r})+\ln A]},
\eqno(3.7)$$
where $K_{\rho}(w)$ is the Macdonald function of order $\rho$.
Vacuum fermion number density (3.7) vanishes at half-integer values
of reduced vortex flux $\F-\Upsilon$ (i.e. at $F={1\over2}$)
as well as at $\cos\Theta=0$.
Otherwise, at large distances from the vortex we get
$$
\N_\x{}_{\stackrel{\ds =}{r\rightarrow
\infty}}-(F-{1\over2}){\sin(F\pi)\over2\pi^2r^2}
\left\{\ba{cc}
(\mu r)^{2F-1}A^{-1}{\ds\Gamma({3\over2}-F)\Gamma({3\over2}-2F)\over\ds
\Gamma(2-F)},& 0<F<{1\over2}\\[0.2cm]
(\mu r)^{1-2F}A{\ds\Gamma(F+{1\over2})\Gamma(2F-{1\over2})\over\ds
\Gamma(1+F)},& {1\over2}<F<1 \\ \ea \right. \,.\eqno(3.8)$$

Integrating Eq.(3.7) over the plane ($x^1,x^2$), we obtain the total
vacuum fermion number
$$\N=-{1\over2}\sgn_0\left[(F-{1\over2})\cos\Theta\right], \eqno(3.9)$$
where
$$\sgn_0(u)=\left\{ \ba{cc}
\sgn(u),& u\neq0\\
0,& u=0\\ \ea \right\}.
$$

\section{Zeta Function}

In the second-quantized theory the operator of energy is defined as

$$
\hat{\E}=\int d^2x\,{1\over2}\bigl[\Psi^+(\x,t), H\Psi(\x,t)\bigr]_- =\si
(E_\lambda a_\lambda^+a_\lambda -E_\lambda b_\lambda^+ b_\lambda
-{1\over2}|E_\lambda|), \eqno(4.1)
$$
thus the vacuum expectation value of the energy takes the form

$$
\E\equiv <\vac|\,\hat{\E}|\vac>=-{1\over2} \si|E_\lambda|=- {1\over2}\int
d^2x\,\tr<\x|\,|H|\,|\x>. \eqno(4.2)$$
The latter expression is ill-defined due to divergences of various
kinds. First, there is a bulk divergence resulting from the integration
over the infinite twodimensional space. But, even if one considers the
vacuum energy density,

$$\E_\x=-{1\over2}\tr<\x|\,|H|\,|\x>, \eqno(4.3)$$
still it remains to be divergent. There is a divergence at large values
of momentum of integration, $k\rightarrow \infty$. To tame this
divergence, let us introduce the zeta function density

$$\zeta_\x(z)=\tr<\x|\,|H|^{-2z}|\x>, \eqno(4.4)
$$
which is ultraviolet convergent at sufficiently large values of $\Re
z$. However, exactly at these values of $\Re z$ the integral
corresponding to Eq.(4.4) is divergent in the infrared region, as
$k\rightarrow 0$. To regularize this last divergence, let us introduce
fermion mass $m$, modifying definition (4.4):
$$
\zeta_\x(z|m)=\tr<\x|\,|\tilde{H}|^{-2z}|\x>, \eqno(4.5)
$$
where
$$
\tilde{H}=-i\ab[\bpar-i\V(\x)] +\beta m, \eqno(4.6)
$$
and it is implied that the complete set of solutions to the equation
$$
\tilde{H}<\x|\lambda>=\tilde{E}_\lambda<\x|\lambda>, \eqno(4.7)
$$
instead of those to Eq.(2.3), is used.

In the background of a singular magnetic vortex (1.1) -- (1.2) the
radial functions of the solutions to Eq.(4.7) take the form:

$$
\left(\ba{c}
\tilde{f}_n\\
\tilde{g}_n\\ \ea
\right) =
{1\over2\sqrt{\pi}}
\left(\ba{c}
\ds{\sqrt{1+m\tilde{E}^{-1}}J_{l-F}(kr)e^{i\chi_s}}\\[0.2cm]
\ds{\sgn(\tilde{E})\sqrt{1-m\tilde{E}^{-1}}J_{l+1-F}(kr)}\\
\ea\right), \qquad l=s(n-n_0)>0, \eqno(4.8)
$$

$$
\left(\ba{c}
\tilde{f}_n\\
\tilde{g}_n\\ \ea
\right) =
{1\over2\sqrt{\pi}}
\left(\ba{c}
\ds{\sqrt{1+m\tilde{E}^{-1}}J_{l'+F}(kr)e^{i\chi_s}}\\[0.2cm]
\ds{-\sgn(\tilde{E})\sqrt{1-m\tilde{E}^{-1}}J_{l'-1+F}(kr)}\\
\ea\right), \qquad l'=s(n_0-n)>0, \eqno(4.9)
$$

$$
\left(\ba{c}
\tilde{f}_{n_0}^{(C)}\\
\tilde{g}_{n_0}^{(C)}\\ \ea
\right) =
{1\over 2\sqrt{\pi[1+\sin(2\tilde{\nu}_{\tilde{E}})\cos(F\pi)}}\times $$
$$\times
\left(\ba{c}
\ds{\sqrt{1+m\tilde{E}^{-1}}[\sin(\tilde{\nu}_{\tilde{E}})J_{-F}(kr)+
\cos(\tilde{\nu}_{\tilde{E}})J_F(kr)]e^{i\chi_s}}\\[0.2cm]
\ds{\sgn(\tilde{E})\sqrt{1-m\tilde{E}^{-1}}[\sin(\tilde{\nu}_{\tilde{E}})
J_{1-F}(kr)-\cos(\tilde{\nu}_{\tilde{E}})J_{-1+F}(kr)]}\\
\ea\right), \eqno(4.10)
$$
where $k=\sqrt{\tilde{E}^2-m^2}$,
$$
\tan(\tilde{\nu}_{\tilde{E}})=\sgn(\tilde{E})\sqrt{{1-m\tilde{E}^{-1}\over
1+m\tE^{-1}}}\left({k\over \mu}\right)^{2F-1}A, \eqno(4.11)
$$
$A$ is given by Eq.(3.5); note that the radial functions of
irregular solution (4.10) satisfy condition (2.14) (see Appendix
A). Note also that Eqs.(4.8) -- (4.10) correspond to the continuum,
$|\tE|>|m|$ \footnote{In a 2+1-, as well as in any odd-, dimensional
space-time mass parameter $m$ in Eq.(4.6) can take both positive
and negative values.}. In addition to them, in the case of
$$
\sgn(m)\cos \Theta<0 ,\eqno(4.12)
$$
an irregular solution corresponding to the bound state appears. Its
radial functions are

$$
\left(\ba{c}
\tilde{f}_{n_0}^{(\BS)}\\
\tilde{g}_{n_0}^{(\BS)}\\ \ea
\right) =
{\kappa\over\pi} \sqrt{{\sin(F\pi)\over 1+(2F-1)m^{-1}E_\BS}}
\left(\ba{c}
\ds{\sqrt{1+m^{-1}E_\BS}K_F(\kappa
r)e^{i\chi_s}}\\[0.2cm]
\ds{\sgn(m)\sqrt{1-m^{-1}E_\BS}
K_{1-F}(\kappa r)]}\\
\ea\right), \eqno(4.13)
$$
where $\kappa=\sqrt{m^2-E_\BS^2}$ and the bound state energy
$\tE=E_\BS$ $(|E_\BS|<|m|)$ is determined implicitly by the equation
$$
{(1+m^{-1}E_\BS)^{1-F}\over (1-m^{-1}E_\BS)^F}
=-\sgn(m)\left({|m|\over\mu}\right)^{2F-1} A. \eqno(4.14)
$$

Regular solutions (4.8) and (4.9) yield the following contribution
to zeta function density (4.5):
$$
[\zeta_\x(z|m)]_\REG={1\over4\pi} \int\limits_0^\infty
dk\,k|\tE|^{-2z}\sum_{\sgn(\tE)} \bigl\{ \sum_{l=1}^\infty
\bigl[(1+m\tE^{-1})J_{l-F}^2(kr)
+(1-m\tE^{-1})J_{l+1-F}^2(kr)\bigr]+$$
$$+\sum_{l'=1}^\infty
\bigl[(1+m\tE^{-1})J_{l'+F}^2(kr)+
(1-m\tE^{-1})J_{l'-1+F}(kr)\bigr]\bigr\}. \eqno(4.15)
$$
Summing over the energy sign and over $l$ and $l'$, we get the
expression
$$
[\zeta_\x(z|m)]_\REG ={1\over\pi} \int\limits_0^\infty dk\,k
|\tE|^{-2z} \int\limits_0^{kr}{dy\over y} \bigl[FJ_F^2(y)+(1-F)
J_{1-F}^2(y)\bigr], \eqno(4.16)
$$
which in the case of $\Re z>1$ is reduced to the form
$$
[\zeta_\x(z|m)]_\REG= {1\over2\pi(z-1)} \int\limits_0^\infty {dk\over
k} |\tE|^{2-2z}\bigl[ FJ_F^2(kr)+(1-F)J_{1-F}^2(kr)\bigr]. \eqno(4.17)
$$

Irregular solution (4.10) yields the following contribution to
Eq.(4.5):
$$
[\zeta_\x(z|m)]_\IRREG ={1\over4\pi} \int\limits_0^\infty
dk\,k|\tE|^{-1-2z}\bigl\{ A\mu^{1-2F}k^{2F}[\tLp-\tLm]J_{-F}^2(kr)+$$
$$+ A\mu^{1-2F}k^{-2(1-F)} [(m-|\tE|)^2\tLp-(m+|\tE|)^2\tLm]
J_{1-F}^2(kr)+$$
$$ +2[(m+|\tE|)\tLp-(m-|\tE|)\tLm] J_{-F}(kr)J_F(kr)+2[(m-|\tE|)\tLp-$$
$$ -(m+|\tE|)\tLm] J_{1-F}(kr)J_{-1+F}(kr)+ A^{-1}\mu^{2F-1}k^{-2F}
[(m+|\tE|)^2\tLp-$$
$$ -(m-|\tE|)^2\tLm] J_F^2(kr)+A^{-1}\mu^{2F-1}k^{2(1-F)} [\tLp-\tLm]
J_{-1+F}^2(kr)\bigr\}, \eqno(4.18)
$$
where summation over the energy sign has been performed and
$$
\tilde{L}_{(\pm)}=[A\mu^{1-2F}k^{-2(1-F)}(-m\pm|\tE|)+
2\cos(F\pi)+A^{-1}\mu^{2F-1}k^{-2F} (m\pm|\tE|)]^{-1}. \eqno(4.19)
$$

The contribution of bound state solution (4.13) to Eq.(4.5) is the
following:
$$
[\zeta_\x(z|m)]_\BS= {\sin(F\pi)\over\pi^2}\,
{\kappa^2|E_\BS|^{-2z}\over m+E_\BS(2F-1)} \bigl[(m+E_\BS)K_F^2(\kappa
r)+(m-E_\BS)K_{1-F}^2(\kappa r)\bigr]. \eqno(4.20)
$$

We show in Appendix C that Eq.(4.17) in the case of $1<\Re z<2$ is
transformed to the following expression
$$
[\zeta_\x(z|m)]_\REG ={|m|^{2(1-z)}\over 2\pi(z-1)} +{\sin(z\pi)\over
\pi^2(z-1)} r^{2(z-1)} \times $$
$$\times \int\limits_{|m|r}^\infty {dw\over
w}(w^2-m^2r^2)^{1-z} \bigl[ FI_F(w)K_F(w)+(1-F)I_{1-F}(w)K_{1-F}(w)\bigr],
\eqno(4.21)
$$
while Eq.(4.18) in the case of ${1\over2}<\Re z<1$ is transformed to
the following one
$$
[\zeta_\x(z|m)]_\IRREG ={\sin(z\pi)\over\pi^2} r^{2(z-1)}
\int\limits_{|m|r}^\infty dw\, w(w^2-m^2r^2)^{-z} \bigl[I_F(w)K_F(w)+
I_{1-F}(w)K_{1-F}(w)\bigr]+$$
$$+{2\sin(F\pi)\over\pi^3}
\sin(z\pi)r^{2(z-1)} \times $$
$$\times \int\limits_{|m|r}^\infty dw\,
w(w^2-m^2r^2)^{-z} { A\mu^{1-2F}({w\over r})^{2F}K_F^2(w)+A^{-1}\mu^{2F-1}
({w\over r})^{2(1-F)}K_{1-F}^2(w)\over  A\mu^{1-2F}({w\over
r})^{2F}+2m+A^{-1}\mu^{2F-1}({w\over r})^{2(1-F)}}-$$
$$-{\sin(F\pi)\over\pi^2}\, {\kappa^2|E_\BS|^{-2z}\over m+E_\BS(2F-1)}
\bigl[(m+E_\BS)K_F^2(\kappa r)+(m-E_\BS)K_{1-F}^2(\kappa r)\bigr];
\eqno(4.22)
$$
here $I_\rho(w)$ is the modified Bessel function of order $\rho$.
The integral in Eq.(4.21) can be analytically continued to domain
${1\over2}<\Re z<2$. In the case of ${1\over2}<\Re z<1$ this integral
is decomposed into two terms:
$$[\zeta_\x(z|m)]_\REG= {|m|^{2(1-z)}\over 2\pi(z-1)}-$$
$$-{\sin(z\pi)\over\pi^2} r^{2(z-1)} \int\limits_{|m|r}^\infty dw\,
w(w^2-m^2r^2)^{-z} \bigl[I_F(w)K_F(w)+I_{1-F}(w)K_{1-F}(w)\bigr]+$$
$$+{2\sin(F\pi)\over\pi^3} \, {\sin(z\pi)\over z-1} r^{2(z-1)}
\int\limits_{|m|r}^\infty dw(w^2-m^2r^2)^{1-z} K_F(w)K_{1-F}(w),
\eqno(4.23)
$$
the last of which can be analytically continued to domain $\Re
z<2$. Note also that the second integral in Eq.(4.22) can be
analytically continued to domain $\Re z<1$.

Summing Eqs.(4.20), (4.22) and (4.23), we get
$$
\zeta_\x(z|m)= {|m|^{2(1-z)}\over 2\pi(z-1)}+ {2\sin(F\pi)\over\pi^3}
\, {\sin(z\pi)\over z-1} r^{2(z-1)} \int\limits_{|m|r}^\infty
dw(w^2-m^2r^2)^{1-z} K_F(w)K_{1-F}(w)+$$
$$+ {2\sin(F\pi)\over\pi^3} \sin(z\pi)r^{2(z-1)}\times$$
$$\times
\int\limits_{|m|r}^\infty dw\, w(w^2-m^2r^2)^{-z}
 {A\mu^{1-2F}({w\over r})^{2F}K_F^2(w)+A^{-1}\mu^{2F-1}({w\over
r})^{2(1-F)} K_{1-F}^2(w)\over A\mu^{1-2F}({w\over
r})^{2F}+2m+A^{-1}\mu^{2F-1}({w\over r})^{2(1-F)}}, \eqno(4.24)
$$
i.e., the terms which are defined only in domain ${1\over2}<\Re
z<1$ are cancelled.

Note that the first term in Eq.(4.24) is identified with the zeta
function density in the noninteracting theory (i.e. in the absence of
any boundary condition and any background field):
$$
\zeta_\x^{(0)}(z|m)={|m|^{2(1-z)}\over 2\pi(z-1)}. \eqno(4.25)
$$
In the noninteracting theory all vacuum values are simply omitted due to
the prescription of normal ordering of the product of operators (see,
for example, \cite{Itzy}). Therefore, one has to subtract
$\zeta_\x^{(0)}$ from $\zeta_\x$ for the reasons of consistency. Doing
this and removing the infrared regulator mass, we obtain the
renormalized zeta function density
$$
\zeta_\x^\ren(z) \equiv \lim_{m\rightarrow 0}
\bigl[\zeta_\x(z|m)-\zeta_\x^{(0)}(z|m)\bigr]=$$
$$={\sin(F\pi)\over\pi^3} \sin(z\pi)r^{2(z-1)} \Bigg\{
{\sqrt{\pi}\over4}\, {\Gamma(1-z)\over \Gamma({3\over2}-z)}
\left[1-2{F(1-F)\over1-z}\right] \Gamma(F-z)\Gamma(1-F-z)+$$
$$+\int\limits_0^\infty dw\, w^{1-2z}\left[K_F^2(w)-K_{1-F}^2(w)\right]
\tanh\left[(2F-1)\ln\left({w\over \mu r}\right)+\ln A\right]\Bigg\}. \eqno(4.26)
$$

An alternative way of regularization of the infrared divergence in
Eq.(4.4) is implemented by means of
$$\zeta_\x(z|M)=\tr<\x|\,|H^2+M^2|^{-z}|\x>, \eqno(4.27)
$$
where $M$ is the parameter of the dimension of mass, and it is implied that
the complete set of solutions to the massless Dirac equation, (2.3) and
(2.5), is used. Regular solutions (2.17) and (2.18) yield the
following contribution to Eq.(4.27):
$$
[\zeta_\x(z|M)]_\REG= {1\over2\pi}\int\limits_0^\infty dk\,
k(k^2+M^2)^{-z} \bigl\{\sum_{l=1}^\infty
\bigg[J_{l-F}^2(kr)+J_{l+1-F}^2(kr)\bigr]+$$
$$
+\sum_{l'=1}^\infty
\bigl[J_{l'+F}^2(kr)+J_{l'-1+F}^2(kr)\bigr]\bigg\}, \eqno(4.28)
$$
where the summation over the energy sign has been performed. Following
literally the same line, as in the derivation of Eq.(4.23), we get
$$
[\zeta_\x(z|M)]_\REG=$$
$$=
{|M|^{2(1-z)}\over2\pi(z-1)}-{\sin(z\pi)\over\pi^2} r^{2(z-1)}
\int\limits_{|M|r}^\infty dw\, w(w^2-M^2r^2)^{-z}
\bigl[I_F(w)K_F(w)+I_{1-F}(w)K_{1-F}(w)\bigr]+$$
$$+{2\sin(F\pi)\over\pi^3}\, {\sin(z\pi)\over z-1}r^{2(z-1)}
\int\limits_{|M|r}^\infty dw(w^2-M^2r^2)^{1-z}K_F(w)K_{1-F}(w).
\eqno(4.29)
$$
Irregular solution (2.19) yields the following contribution to
Eq.(4.27):
$$
[\zeta_\x(z|M)]_\IRREG={1\over4\pi} \int\limits_0^\infty dk\,
k(k^2+M^2)^{-z}\Bigg\{ A\left({k\over\mu}\right)^{2F-1} \left[\Lp-\Lm\right] 
\left[J_{-F}^2(kr)+J_{1-F}^2(kr)\right]+$$
$$+2\left[\Lp+\Lm\right]\left[J_{-F}(kr)J_F(kr)-J_{1-F}(kr)J_{-1+F}(kr)\right]+$$
$$+A^{-1}\left({k\over\mu}\right)^{1-2F}\left[\Lp-\Lm\right]
\left[J_F^2(kr)+J_{-1+F}^2(kr)\right]\Bigg\},
\eqno(4.30)
$$
where the summation over the energy sign has been performed and
$L_{(\pm)}$ is given by Eq.(3.6). We show in Appendix D that Eq.(4.30)
at ${1\over2}<\Re z<1$ is transformed to the following expression
$$
[\zeta_\x(z|M)]_\IRREG={\sin(z\pi)\over\pi^2} r^{2(z-1)}
\int\limits_{|M|r}^\infty dw\, w(w^2-M^2r^2)^{-z} \bigl[I_F(w)K_F(w)+
I_{1-F}(w)K_{1-F}(w)\bigr]+$$
$$+{\sin(F\pi)\over\pi^3}
\sin(z\pi)r^{2(z-1)} \int\limits_{|M|r}^\infty dw\, w(w^2-M^2r^2)^{-z}
\bigl\{ K_F^2(w)+K_{1-F}^2(w)+$$
$$+\bigl[ K_F^2(w)-K_{1-F}^2(w)\bigr]
\tanh\bigl[(2F-1)\ln({w\over \mu r})+\ln A\bigr]\bigr\}. \eqno(4.31)
$$
Summing Eqs.(4.29) and (4.31), we get the expression
$$
\zeta_\x(z|M)={|M|^{2(1-z)}\over2\pi(z-1)} +{2\sin(F\pi)\over\pi^3}\,
{\sin(z\pi)\over z-1} r^{2(z-1)} \int\limits_{|M|r}^\infty dw\,
(w^2-M^2r^2)^{1-z}K_F(w)K_{1-F}(w)+$$
$$+{\sin(F\pi)\over\pi^3} \sin(z\pi) r^{2(z-1)} \int\limits_{|M|r}^\infty
dw\, w(w^2-M^2r^2)^{-z}\bigl\{ K_F^2(w)+K_{1-F}^2(w)+$$
$$+\bigl[ K_F^2(w)-K_{1-F}^2(w)\bigr]\tanh\bigl[ (2F-1)\ln({w\over\mu
r})+\ln A\bigr]\bigr\}, \eqno(4.32)
$$
which is analytically continued to domain $\Re z<1$.

One can see that the first term in Eq.(4.32) corresponding to the zeta
function density in the noninteracting theory coincides with Eq.(4.25)
under evident substitution $m\rightarrow M$, whereas the last integrals
in Eqs.(4.32) and (4.24) differ essentially. However, the renormalized
zeta function density
$$
\zeta_\x^\ren(z) \equiv \lim_{M\rightarrow 0}
[\zeta_\x(z|M)-\zeta_\x^{(0)}(z|M)] \eqno(4.33)
$$
coincides with that given by Eq.(4.26).

Although the former way of regularization of the infrared divergence,
resulting in Eq.(4.24), looks much more consistent, the latter one,
resulting in Eq.(4.32), appears to be technically simpler. Since both
of them lead to the same results upon the removal of infrared
regulators, we shall adopt, just for brevity, the latter one in the rest
of the present paper. The only exception will be in the next section
where the use of fermion mass $m$ as an infrared regulator for the
effective action in 2+1-dimensional space-time is generically
inevitable.

\section{Energy Density and Effective Potential}

Recalling the formal expressions for the vacuum energy and zeta
function densities, Eqs.(4.3) and (4.4), one can easily deduce that the
physical (renormalized) vacuum energy density is expressed through the
renormalized zeta function density at $z=-{1\over2}$:
$$
\E_\x^\ren=-{1\over2}\zeta_\x^\ren(-{1\over2}). \eqno(5.1)
$$
In the background of a singular magnetic vortex (1.1) -- (1.2), using
Eq.(4.26), we get the expression
$$
\E_\x^\ren={\sin(F\pi)\over2\pi r^3} \biggl\{
{{1\over2}-F\over6\cos(F\pi)}\left[{3\over4}-F(1-F)\right]+$$
$$+{1\over\pi^2}
\int\limits_0^\infty dw\, w^2
\bigl[K_F^2(w)-K_{1-F}^2(w)\bigr]\tanh\bigl[
(2F-1)\ln({w\over\mu r})+\ln A\bigr]\biggr\}. \eqno(5.2)
$$
At noninteger values of reduced vortex flux $\F-\Upsilon$ (i.e. 
at $0<F<1$) vacuum energy density (5.2) is positive. At half-integer 
values of the reduced vortex flux ($F={1\over2}$) we get
$$
\E_\x^\ren\big|_{F={1\over2}} ={1\over24\pi^2 r^3}. \eqno(5.3)
$$
In the case of $\cos\,\Theta=0$ we get
$$
\E_\x^\ren={\tan(F\pi)\over4\pi r^3} \bigl(F-{1\over2}\bigr)
\left[{1\over3}F(1-F)-{1\over4}\mp{1\over2}\bigl(F-{1\over2}\bigr)
\right], \qquad \Theta=\pm s{\pi\over2}. \eqno(5.4)
$$
If $\cos\,\Theta\neq0$, then at large distances from the vortex we get
$$
\E_\x^\ren \equals_{r\rightarrow \infty} {\tan(F\pi)\over4\pi
r^3} \bigl(F-{1\over2}\bigr)
\left[{1\over3}F(1-F)-{1\over4}+{1\over2}|F-{1\over2}|\right].
\eqno(5.5)
$$

Going over to imaginary time $t=-i\tau$,  let us consider the
effective action in 2+1-dimensional Euclidean space-time:
$$
S_{(2+1)}^\eff[\V(\x)] =-\ln\bigl\{ N^{-1}\int d\Psi\,d\Psi^+\exp[-\int
d\tau\,d^2x\,\Psi^+(-i\beta\partial_\tau-
i\beta \tilde{H})\Psi]\bigr\}=$$
$$=-\ln\Det\bigl[(-i\beta\partial_\tau-
i\beta\tilde{H})\tilde{m}^{-1}\bigr]; \eqno(5.6)
$$
here $N$ is a normalization factor, parameter $\tilde{m}$ is
inserted just for the dimension reasons, while fermion mass $m$ (see
Eq.(4.6)) is introduced in order to tame the infrared divergence. The
real part of the effective action is presented in the form \footnote{The
imaginary part of the effective action vanishes in the case of a static 
background.}
$$
\Re S_{(2+1)}^\eff[\V(\x)]=-{1\over2}\int d\tau
d^2x\,\tr<\x,\tau|\ln[(-\partial_\tau^2+\tilde{H}^2)\tilde{m}^{-2}]|\x,\tau>.
\eqno(5.7)
$$
Let us define the zeta function density in threedimensional space
($x^1,x^2,\tau$):
$$
\zeta_{\x,\tau}(z|m)=\tr<\x,\tau|\,(-\partial_\tau^2 +
\tilde{H}^2)^{-z}|\x,\tau>. \eqno(5.8)
$$
Then an ultraviolet regularization of Eq.(5.7) can be achieved by
expressing its integrand through Eq.(5.8):
$$
-{1\over2}\tr<\x,\tau|\ln[(-\partial_\tau^2+\tilde{H}^2)\tilde{m}^{-2}]
|\x,\tau>= {1\over2}\bigl[{d\over
dz}\zeta_{\x,\tau}(z|m)\bigr]\big|_{z=0} +
{1\over2}\zeta_{\x,\tau}(0|m)\ln\tilde{m}^2. \eqno(5.9)
$$

In the background of a singular magnetic vortex (1.1) -- (1.2) we
get, similarly to Eq.(4.24), the following expression
$$
\zeta_{\x,\tau}(z|m)={|m|^{3-2z}\over4\pi^{3\over2}}\,
{\Gamma(z-{3\over2})\over\Gamma(z)} -$$
$$-{\sin(F\pi)\over\pi^{7\over2}}
\cos(z\pi){\Gamma(z-{3\over2})\over\Gamma(z)}r^{2z-3}
\int\limits_{|m|r}^\infty dw(w^2-m^2r^2)^{{3\over2}-z}
 K_F(w)K_{1-F}(w)-$$
$$-{\sin(F\pi)\over\pi^{7\over2}}\cos(z\pi)
{\Gamma(z-{1\over2})\over\Gamma(z)} r^{2z-3}\times$$
$$\times\int\limits_{|m|r}^\infty
dw\, w(w^2-m^2r^2)^{{1\over2}-z}
 {A\mu^{1-2F}({w\over r})^{2F}K_F^2(w)+A^{-1}\mu^{2F-1}({w\over
r})^{2(1-F)}K_{1-F}^2(w)\over A\mu^{1-2F}({w\over r})^{2F}+2m+A^{-1}
\mu^{2F-1}({w\over r})^{2(1-F)}}; \eqno(5.10)
$$
note that the first term in Eq.(5.10) corresponds to the case of the
noninteracting theory:
$$
\zeta_{\x,\tau}^{(0)}(z|m)={|m|^{3-2z}\over4\pi^{3\over2}}\,
{\Gamma(z-{3\over2})\over\Gamma(z)}. \eqno(5.11)
$$
Note also relation
$$
\zeta_{\x,\tau}(0|m)=0, \eqno(5.12)
$$
which ensures the independence of the effective action on
$\tilde{m}^2$; thus Eq.(5.9) takes the form
$$
-{1\over2}\tr\langle
\x,\tau|\ln(-\partial_\tau^2+\tilde{H}^2)\tilde{m}^{-2}|\x,\tau\rangle=
{|m|^3\over6\pi}-
 {2\sin(F\pi)\over3\pi^3 r^3}\, \int\limits_{|m|r}^\infty
dw(w^2-m^2r^2)^{3\over2}K_F(w)K_{1-F}(w)+$$
$$+{\sin(F\pi)\over \pi^3r^3}
\int\limits_{|m|r}^\infty dw\, w(w^2-m^2r^2)^{1\over2}
 {A\mu^{1-2F}({w\over r})^{2F}K_F^2(w)+A^{-1}
\mu^{2F-1}({w\over r})^{2(1-F)}K_{1-F}^2(w)\over A\mu^{1-2F}({w\over
r})^{2F}+2m+A^{-1}\mu^{2F-1}({w\over r})^{2(1-F)}}. \eqno(5.13)
$$

The effective potential in the massless theory is defined as
$$
{\cal U}^\eff(\x,\tau)=- {1\over2}\lim_{m\rightarrow 0}\tr
\langle\x,\tau|\ln \biggl[
{(-\partial_\tau^2+\tilde{H}^2)\tilde{m}^{-2}\over
(-\partial_\tau^2-\bpar^2+m^2)\tilde{m}^{-2}}\biggr]|\x,\tau\rangle.
\eqno(5.14)
$$
Defining the renormalized zeta function density
$$
\zeta_{\x,\tau}^\ren(z)=\lim_{m\rightarrow 0}
\bigl[\zeta_{\x,\tau}(z|m)- \zeta_{\x,\tau}^{(0)}(z|m)\bigr],
\eqno(5.15)
$$
we get, similarly to Eq.(5.9),
$$
{\cal U}^\eff(\x,\tau)={1\over2} \bigl[{d\over
dz}\zeta_{\x,\tau}^\ren(z)\bigr]\mid_{z=0}
+{1\over2}\zeta_{\x,\tau}^\ren(0)\ln\tilde{m}^2. \eqno(5.16)
$$
In the background of a singular magnetic vortex (1.1) -- (1.2), using
Eq.(5.10), we get
$$
\zeta_{\x,\tau}^\ren(z)=-{\sin(F\pi)\over 2\pi^{7\over2}}\cos(z\pi)
{\Gamma(z-{1\over2})\over\Gamma(z)} r^{2z-3}\times$$
$$\times
\left\{{\sqrt{\pi}\over4}\, {\Gamma({3\over2}-z)\over\Gamma(2-z)}
\left[1-{4F(1-F)\over 3-2z}\right] \Gamma\bigl({1\over2}-z+F\bigr)
\Gamma\bigl({3\over2}-z-F\bigr)+\right.$$
$$+\left.\int\limits_0^\infty dw\, w^{2(1-z)}
\bigl[K_F^2(w)-K_{1-F}^2(w)\bigr] \tanh\bigl[(2F-1)\ln({w\over\mu
r})+\ln A\bigr]\right\}, \eqno(5.17)
$$
in particular
$$\zeta_{\x,\tau}^\ren(0)=0, \eqno(5.18)
$$
and, thence, we arrive at the remarkable relation
$$
{\cal U}^\eff(\x,\tau)=\E_\x^\ren, \eqno(5.19)
$$
where $\E_\x^\ren$ is given by Eq.(5.2).

Although the last relation looks rather natural and even evident, let
us emphasize here that it is a consequence of the relation between the
renormalized zeta function densities of different spatial dimensions,
$$
\bigl[{d\over
dz}\zeta_{\x,\tau}^\ren(z)\bigr]\big|_{z=0}=-\zeta_\x^\ren(-{1\over2}),
\eqno(5.20)
$$
and relation (5.18). As it has been shown in Ref. \cite{SitR98},
relation (5.20) can be in general broken in spaces of higher
dimensions. Moreover, both left- and right-hand sides of Eq.(5.20) can
have nothing to do with the true vacuum energy density. Fortunately,
this is not relevant for the case considered in the present paper,
and, indeed, in the background of a singular magnetic vortex the vacuum
energy density coincides with the effective potential.

\section{Current}

Let us regard the 2-dimensional space $(x^1,x^2)$ as a 1+1-dimensional 
space-time with the Wick-rotated time axis. The
Clifford algebra in this space-time has the exact irreducible
representation with the above $\alpha$ matrices playing now the role of
the $\gamma$ matrices and the $\beta$ matrix playing the role similar
to that of the $\gamma^5$ matrix in a 3+1-dimensional space-time. Introducing
mass parameter $M$ to tame the infrared divergence, one can define
the trace of two-point causal Green's function with the $\alpha$
matrix inserted between the field operators:
$$
\J(\x,\x'|M)=<\vac|T\Psi^+(\x',0)\ab\Psi(\x,0)|\vac>=
\tr<\x|\,\ab(H-iM)^{-1}|\x'>, \eqno(6.1)$$
where $T$ is the symbol of time ordering in the 1+1-dimensional space-time.
Inserting damping factor $(E^2+M^2)^{-z}$ (where $\Re\, z>0$) into
the integral corresponding to Eq.(6.1), we define matrix element
$$
\J(\x,\x';z|M)=\tr<\x|\,\ab(H+iM)(H^2+M^2)^{-1-z}|\x'>. \eqno(6.2)$$
Returning to the massless theory in the 2+1-dimensional space-time
($x^1,x^2,t)$, let us define the vacuum current in the conventional way
(compare with Eqs. (3.1) -- (3.3))
$$\j(\x)=<\vac|{1\over2}\left[\Psi^+(\x,t),\ab\,\Psi(\x,t)\right]_-|\vac>=
-{1\over2}\tr<\x|\,\ab\,\sgn(H)|\x>. \eqno(6.3)$$
One can notice relation
$$\j(\x)=-{1\over2}\J(\x,\x;-{1\over2}|0), \eqno(6.4)$$
so that in the following the matrix element (6.2) in the coincidence
limit $\x'=\x$ will be regarded as a generalized current.

In the background of a singular magnetic vortex (1.1) -- (1.2)
radial component
$$
J_r(\x,\x;z|M)=r^{-1}[x^1J_1(\x,\x;z|M)+x^2J_2(\x,\x;z|M)] \eqno(6.5)$$
vanishes, whereas angular component
$$J_\varphi(\x,\x;z|M)=r^{-1}[x^1J_2(\x,\x;z|M)-x^2J_1(\x,\x;z|M)]
\eqno(6.6)$$
is nonvanishing. The contribution of regular solutions (2.17) and
(2.18) to Eq.(6.6) is given by the expression
$$[J_\varphi(\x,\x;z|M)]_\REG={s\over\pi}\int\limits_0^\infty
dk{k^2\over (k^2+M^2)^{1+z}} \biggl[\sum_{l=1}^\infty
J_{l-F}(kr)J_{l+1-F}(kr) -$$
$$-
\sum_{l'=1}^\infty
J_{l'+F}(kr)J_{l'-1+F}(kr)\biggr].  \eqno(6.7)$$
Performing the summation over $l$ and $l'$, we get
$$[J_\varphi(\x,\x;z|M)]_\REG= {s\over\pi}\int\limits_0^\infty dk
{k^2(k^2+M^2)^{-1-z}}\biggl\{ FJ_F(kr)J_{-1+F}(kr)-$$
$$-(1-F)J_{1-F}(kr)J_{-F}(kr)-{1\over2}kr
[J_F^2(kr)+J_{-1+F}^2(kr)-J_{-F}^2(kr)-J_{1-F}^2(kr)]\biggr\}.
\eqno(6.8)$$
Transforming the integral in the last expression similarly to
that as in derivation of $[\zeta_\x(z|M)]_\REG$ (4.29) in Section IV,
we get
$$
[J_\varphi(\x,\x;z|M)]_\REG={s\sin(z\pi)\over\pi^2}r^{2z-1}
\int\limits_{|M|r}^\infty dw{w^2(w^2-M^2r^2)^{-1-z}}\times$$
$$\times\biggl\{
I_F(w)K_{1-F}(w)-I_{1-F}(w)K_F(w)+$$
$$+{2\sin(F\pi)\over\pi}
[wK_F^2(w)-wK_{1-F}^2(w)
-(2F-1)K_F(w)K_{1-F}(w)]\biggr\}. \eqno(6.9)$$ 
The contribution of irregular solution (2.19) to Eq.(6.6) is given
by expression
$$
[J_\varphi(\x,\x;z|M)]_\IRREG={s\over2\pi}\int\limits_0^\infty
dk{k(k^2+M^2)^{-1-z}}\times$$
$$\times \biggl\{
A\biggl({k\over\mu}\biggr)^{2F-1}[(k+iM)\Lp-(k-iM)\Lm]J_{-F}(kr)J_{1-F}(kr)+$$
$$+ [(k+iM)\Lp+(k-iM)\Lm][J_F(kr)J_{1-F}(kr)-
J_{-F}(kr)J_{-1+F}(kr)]-$$
$$-A^{-1}\biggl({k\over\mu}\biggr)^{1-2F}[(k+iM)\Lp-(k-iM)\Lm]J_F(kr)J_{-1+F}(kr)\biggr\},
\eqno(6.10)$$
where $A$ and $L_{(\pm)}$ are given by Eqs.(3.5) and (3.6),
respectively. Transforming the integral in Eq.(6.10) similarly to
that as in Appendix D, we get
$$
[J_\varphi(\x,\x;z|M)]_\IRREG={s\sin(z\pi)\over\pi^2}r^{2z-1}
\int\limits_{|M|r}^\infty dw{w^2(w^2-M^2r^2)^{-1-z}}\times$$
$$\times \biggl(I_{1-F}(w)K_F(w)-I_F(w)K_{1-F}(w)- {2\sin(F\pi)\over\pi}
K_F(w)K_{1-F}(w)\times$$
$$\times \bigl\{\tanh[(2F-1)\ln({w\over\mu r})+\ln A]- {iMr\over
w\cosh[(2F-1)\ln({w\over\mu r})+\ln A]}\bigr\}\biggr). \eqno(6.11)$$
Summing Eqs.(6.9) and (6.11), we get
$$
J_\varphi(\x,\x;z|M)={2s\sin(F\pi)\over\pi^3}\sin(z\pi)r^{2z-1}
\int\limits_{|M|r}^\infty dw\,{w^2(w^2-M^2r^2)^{-1-z}}\times$$
$$\times \biggl( w[K_F^2(w)-K_{1-F}^2(w)]-K_F(w)K_{1-F}(w)
\bigl\{2F-1+$$
$$+\tanh[(2F-1)\ln\bigl({w\over\mu r}\bigr)+\ln A]-{iMr\over
w\cosh[(2F-1)\ln({w\over\mu r})+\ln A]}\bigr\}\biggr). \eqno(6.12) $$
Note that Eqs.(6.9) and (6.11) are valid at $-{1\over2}<\Re z<0$, and
their sum, Eq.(6.12), can be analytically continued to half-plane
$\Re z<0$. This may look somewhat embarrassing, since the initial
motivation, as presented above, was to consider the case of $\Re z>0$.
However, the situation can be cured by means of analytic continuation
using partial integration. Namely, integrating Eq.(6.12) by parts, we
get the expression 
$$J_\varphi(\x,\x;z|M)=-{s\sin(F\pi)\over\pi^3}\,
{\sin(z\pi)\over z}r^{2z-1}\int\limits_{|M|r}^\infty
dw(w^2-M^2r^2)^{-z}\times$$
$$\times\biggl[w\bigl[K_F^2(w)-K_{1-F}^2(w)\bigr]+\bigl[K_F^2(w)+K_{1-F}^2(w)
\bigr]\bigl\{ w\tanh\bigl[(2F-1)\ln\bigl({w\over\mu r}\bigr)+\ln A
\bigr]-$$
$$-{iMr\over\cosh[(2F-1)\ln({w\over\mu r})+\ln A]}\bigr\}-
{K_F(w)K_{1-F}(w)\over \cosh[(2F-1)\ln({w\over\mu r})+\ln A]}\times$$
$$\times \biggl( {2F-1\over\cosh[(2F-1)\ln({w\over\mu r})+\ln
A]}+{iMr\over w}\bigl\{(2F-1)\tanh[(2F-1)\ln\bigl({w\over\mu r}\bigr)+
\ln A\bigr]+1\bigr\}\biggr)\biggr], \eqno(6.13)$$
which is analytically continued to domain $\Re z<1$. Integrating
Eq.(6.12) by parts $N$ times, one can get the expression for the
generalized current which is analytically continued to domain 
$\Re z<N$.

Taking into account Eq.(6.4), we get the following expression for the
vacuum current:
$$j_\varphi(\x)={s\sin(F\pi)\over\pi r^2}\biggl\{
{(F-{1\over2})^2\over4\cos(F\pi)}-{1\over\pi^2}\int\limits_0^\infty dw\,
wK_F(w)K_{1-F}(w) \tanh\bigl[[(2F-1)\ln\bigl({w\over\mu r}\bigr)
+\ln A\bigr]\biggr\}; \eqno(6.14)$$
recall that radial component $j_r(\x)$ is vanishing, see Eq.(6.5).
At $\cos\Theta=0$ we get
$$j_\varphi(\x)={s\tan(F\pi)\over4\pi
r^2}(F-{1\over2})(F-{1\over2}\pm1), \qquad \Theta=\pm s{\pi\over2}.
\eqno(6.15)$$
At half-integer values of the reduced vortex flux $(F={1\over2})$,
taking into account relation
$$A|_{F={1\over2}}=\tan\bigl(s{\Theta\over2}+{\pi\over4}\bigr),
\eqno(6.16)$$
we get
$$j_\varphi(\x)|_{F={1\over2}}=-{\sin\Theta\over4\pi^2r^2}.
\eqno(6.17)$$
If $\cos\Theta\neq0$ and $F\neq{1\over2}$, then at large distances from
the vortex we get 
$$j_\varphi(\x){}_{\stackrel{\ds =}{r\rightarrow
\infty}}\, {s\tan(F\pi)\over4\pi r^2}|F-{1\over2}|\left(|F-{1\over2}|-1\right).
\eqno(6.18)$$

\section{Parity Breaking Condensate}

Since twodimensional massless Dirac Hamiltonian (2.5) anticommutes
with the $\beta$ matrix
$$[H,\beta]_+=0, \eqno(7.1)$$
the Dirac equation (2.3) is invariant under the parity transformation
$$E_\lambda\rightarrow -E_\lambda, \qquad <\x|\lambda>\rightarrow
\,\beta<\x|\lambda>. \eqno(7.2)$$
However, this invariance is violated by boundary condition (2.14), 
unless $\cos\Theta=0$. Consequently, the parity breaking
condensate emerges in the vacuum:
$$\C_\x=<\vac|{1\over2}[\Psi^+(\x,t),\beta\,\Psi(\x,t)]_-|\vac>=
-{1\over2}\tr<\x|\beta\,\sgn(H)|\x>. \eqno(7.3)$$

Let us start with the regularized condensate
$$\C_\x(z|M)=-{1\over2}\tr<\x|\beta\, H(H^2+M^2)^{-{1\over2}-z}|\x>.
\eqno(7.4)$$
The contribution of regular solutions (2.17) and (2.18) to Eq.(7.4) 
is cancelled upon summation over the sign of energy. Thus, only
the contribution of irregular solution (2.19) to Eq.(7.4) survives:
$$
\C_\x(z|M)=-{1\over8\pi}\int\limits_0^\infty {dk\,
k^2(k^2+M^2)^{-{1\over2}-z}}\biggl\{
A\biggl({k\over\mu}\biggr)^{2F-1}\bigl[\Lp+\Lm\bigr]
\bigl[J_{-F}^2(kr) -J_{1-F}^2(kr)\bigr]+$$
$$+2\bigl[\Lp-\Lm\bigr]
\bigl[J_{-F}(kr)J_F(kr)+J_{1-F}(kr)J_{-1+F}(kr)\bigr]+$$
$$
+A^{-1}\biggl({k\over\mu}\biggr)^{1-2F}\bigl[\Lp+\Lm\bigr]
\bigl[J_F^2(kr)-J_{-1+F}^2(kr)\bigr]\biggr\}, \eqno(7.5)$$
where $A$ and $L_{(\pm)}$ are given by Eqs.(3.5) and (3.6),
respectively. Transforming the integral in Eq.(7.5) similarly
to that as in Appendix D, we get
$$
\C_\x(z|M)=-{\sin(F\pi)\over2\pi^3}\cos(z\pi)r^{2(z-1)}
\int\limits_{|M|r}^\infty dw\, w^2(w^2-M^2r^2)^{-{1\over2}-z}
 {K_F^2(w)+K_{1-F}^2(w)\over \cosh[(2F-1)\ln({w\over\mu
r})+\ln A]}, \eqno(7.6)$$
where $\Re z<{1\over2}$. Then vacuum condensate (7.3) is given 
by the following expression:
$$\C_\x\equiv \C_\x(0|0)=- {\sin(F\pi)\over2\pi^3r^2}\int\limits_0^\infty dw\,
w{K_F^2(w)+K_{1-F}^2(w)\over\cosh[(2F-1)\ln({w\over\mu r})+\ln A]}.
\eqno(7.7)$$
Evidently, Eq.(7.7) vanishes at $\cos\Theta=0$. At half-integer values
of the reduced vortex flux $(F={1\over2})$ we get
$$\C_\x\big|_{F={1\over2}}=-{\cos\Theta\over 4\pi^2 r^2}. \eqno(7.8)$$
If $\cos\Theta\neq0$ and $F\neq{1\over2}$, then at large 
distances from the vortex we get
$$
\C_\x{}_{\stackrel{\ds =}{r\rightarrow
\infty}}-{\sin(F\pi)\over2\pi^2r^2} \left\{\ba{cc} (\mu
r)^{2F-1}A^{-1}{\ds\Gamma({3\over2}-F)\Gamma({3\over2}-2F)\over \ds
\Gamma(1-F)}, & 0<F<{1\over2}\\[0.2cm]
(\mu r)^{1-2F}A {\ds\Gamma(F+{1\over2})\Gamma(2F-{1\over2})\over\ds
\Gamma(F)},& {1\over2}<F<1\\ \ea \right. \,.\eqno(7.9)$$
Integrating Eq.(7.7) over the plane ($x^1,x^2$), we obtain the total
vacuum condensate
$$
\C\equiv\int d^2x\,\C_\x=-{\sgn_0(\cos\Theta)\over4|F-{1\over2}|}.
\eqno(7.10)$$
Thus, the total vacuum condensate is infinite at $F={1\over2}$ if
$\cos\Theta\neq0$.

\section{Absence of Parity Anomaly}

Let us define the generalized twodimensional axial current
$$\J^3(\x,\x;z|M)=i\tr<\x|\ab\beta(H+iM)(H^2+M^2)^{-1-z}|\x>.
\eqno(8.1)$$
Owing to the twodimensionality, the components of current (8.1) are
related to the components of the generalized current considered in
Section VI (Eq.(6.2) at $\x'=\x$)
$$J_r^3=sJ_\varphi, \qquad J^3_\varphi=-sJ_r. \eqno(8.2)$$
In the background of a singular magnetic vortex (1.1) -- (1.2) the
divergence of the current $\J$ is vanishing (see Eq.(6.12)):
$$\bpar\cdot\J(\x,\x;z|M)=0, \eqno(8.3)$$
whereas the divergence of the current $\J^3$ is nonvanishing, owing to
the relation
$$\bpar\cdot\J^3(\x,\x;z|M)=s\,\bpar\times\J(\x,\x;z|M). \eqno(8.4)$$

Let us consider the effective action in 1+1-dimensional Euclidean 
space-time
$$S^{\rm eff}_{(1+1)}[\V(\x)]=-\int d^2x\,\tr<\x|
\ln(H\tilde{M}^{-1})|\x>, \eqno(8.5)$$
where $\tilde{M}$ is the parameter of the dimension of mass. The invariance
of action (8.5) under the gauge transformation,
$$
V(\x)\rightarrow \V(\x)+\bpar\,\Lambda(\x),$$
$$<\x|\rightarrow e^{i\Lambda(\x)}<\x|, \qquad |\x>\rightarrow
e^{-i\Lambda(\x)}|\x>, \eqno(8.6)$$
is stipulated by conservation law
$$\lim_{M\rightarrow 0\atop z\rightarrow 0}\bpar\cdot\J(\x,\x;z|M)=0.
\eqno(8.7)$$
If conservation law
$$\lim_{M\rightarrow 0\atop z\rightarrow 0}\bpar\cdot\J^3(\x,\x;z|M)=0
\eqno(8.8)$$
holds, then action (8.5) is invariant under the localized version
of the parity transformation (compare with Eq.(7.2))
$$
\V(\x)\rightarrow \V(\x)+\bpar\,\beta\Lambda(\x),$$
$$<\x|\rightarrow e^{i\beta\Lambda(\x)}<\x|, \qquad |\x>\rightarrow
|\x>e^{i\beta\Lambda(\x)}. \eqno(8.9)$$
The breakdown of the latter symmetry,
$$\lim_{M\rightarrow 0\atop z\rightarrow
0}\bpar\cdot\J^3(\x,\x;z|M)\neq0, \eqno(8.10)$$
is denoted as a parity anomaly, i.e., the axial anomaly in
1+1-dimensional space-time.

Note that in classical theory both gauge and localized parity
symmetries are conserved. This is reflected by the formal invariance of
action (8.5) under both transformations (8.6) and (8.9). However, in
quantum theory, in order to calculate the effective action in a certain
background, one has to use regularization which in fact breaks the
localized parity symmetry. Namely, one has to substitute
$\ln[(H-iM)\tilde{M}^{-1}]$ for $\ln(H\tilde{M}^{-1})$ into Eq.(8.5) 
(compare with Eqs.(5.6)-(5.7)), where regulator mass $M$ is 
the symmetry breaking parameter. That
is why generalized currents (6.2) at $\x'=\x$ and (8.1) come into
play. The divergence of the latter current can be presented in the form
$$
\bpar\cdot\J^3(\x,\x;z|M)=
2\tilde{\zeta}_\x(z|M)-2M^2\tilde{\zeta}_\x(z+1|M)- 
4iM\C_\x(z+{1\over2}|M), \eqno(8.11)$$
where $\C_\x(z|M)$ is regularized condensate (7.4) and
$$\tilde{\zeta}_\x(z|M)=\tr<\x|\beta\,(H^2+M^2)^{-z}|\x> \eqno(8.12)$$
is the modified (by insertion of the $\beta$ matrix) zeta function
density.

In the background of a singular magnetic vortex, conservation law
(8.7) holds, as a consequence of Eq.(8.3). Incidentally, the
regularized condensate is given by Eq.(7.6). As to the
modified zeta function density, we get the following 
expression for the contribution of regular solutions (2.17) and (2.18)
to Eq.(8.12):
$$
[\tilde{\zeta}_\x(z|M)]_{\REG}=
\frac{1}{2\pi}\int_0^{\infty}dk k (k^2+M^2)^{-z}
\Bigg\{ \sum_{l=1}^{\infty} \left[ J_{l-F}^2(kr)-J_{l+1-F}^2(kr) \right]+$$
$$+
\sum_{l'=1}^{\infty} \left[ J_{l'+F}^2(kr)-J_{l'-1+F}^2(kr) \right] \Bigg\},
\eqno(8.13)$$
which, upon summation over $l$ and $l'$, takes the form
$$
[\tilde{\zeta}_\x(z|M)]_{\REG}=
\frac{1}{2\pi}\int_0^{\infty}dk k (k^2+M^2)^{-z}
\Big[ J_{1-F}^2(kr)-J_{F}^2(kr) \Big] .
\eqno(8.14)$$
Irregular solution (2.19) yields the following contribution to Eq.(8.12):
$$
[\tilde{\zeta}_\x(z|M)]_{\IRREG}=
\frac{1}{4\pi}\int_0^{\infty}dk k (k^2+M^2)^{-z}
\Bigg\{ A \left(\frac{k}{\mu} \right)^{2F-1}
\left[ L_{(+)}-L_{(-)} \right]\times$$
$$\times \left[ J_{-F}^2(kr)-J_{1-F}^2(kr) \right]+
2 \left[ L_{(+)}+L_{(-)} \right]
\left[ J_{-F}(kr)J_{F}(kr)+J_{1-F}(kr)J_{-1+F}(kr) \right]+$$
$$+
 A^{-1} \left(\frac{k}{\mu} \right)^{1-2F}
\left[ L_{(+)}-L_{(-)} \right] \left[ J_{F}^2(kr)-J_{-1+F}^2(kr) \right] \Bigg\} ,
\eqno(8.15)$$
where $A$ and $L_{(\pm)}$ are given by Eqs.(3.5) and (3.6). Similarly to 
the case of zeta function density ${\zeta}_\x(z|M)$ (4.27)
(see Section IV and Appendix D), we recast Eqs.(8.14) and (8.15) for 
$\frac{1}{2} < \Re\, z < 1$ into the following form 
$$
[\tilde{\zeta}_\x(z|M)]_{\REG}=
{\sin(z\pi)\over\pi^2}
r^{2(z-1)}\int\limits_{|M|r}^\infty dw\, w(w^2-M^2r^2)^{-z}
\Big[ I_{1-F}(w)K_{1-F}(w)-I_{F}(w)K_{F}(w)\Big]
\eqno(8.16)$$
and
$$
[\tilde{\zeta}_\x(z|M)]_{\IRREG}=
{\sin(z\pi)\over\pi^2}
r^{2(z-1)}\int\limits_{|M|r}^\infty dw\, w(w^2-M^2r^2)^{-z}
\Big[ I_{F}(w)K_{F}(w)-I_{1-F}(w)K_{1-F}(w)\Big]+$$
$$
+
{\sin(F\pi)\over\pi^3}
\sin(z\pi)r^{2(z-1)} \int\limits_{|M|r}^\infty dw\, w(w^2-M^2r^2)^{-z}
\Bigg\{ K_{F}^2(w)-K_{1-F}^2(w)+$$
$$
+
\Big[K_{F}^2(w)+K_{1-F}^2(w)\Big]
 \tanh \Big[ (2F-1)\ln \left(\frac{w}{\mu r}\right)
+\ln A \Big] \Bigg\},
\eqno(8.17)$$
resulting in the expression
$$
\tilde{\zeta}_\x(z|M)={\sin(F\pi)\over\pi^3}
\sin(z\pi)r^{2(z-1)}\int\limits_{|M|r}^\infty dw\, w(w^2-M^2r^2)^{-z}\times$$
$$\times \biggl\{K_F^2(w)-K_{1-F}^2(w)+
\bigl[K_F^2(w)+K_{1-F}^2(w)\bigr]\tanh\bigl[(2F-1)\ln\bigl({w\over \mu
r}\bigr)+\ln A\bigr]\biggr\}, \eqno(8.18)$$
which is analytically continued to domain $\Re z<1$.

 Extending the domain of definition in $z$ in Eqs.(7.6)
and (8.18) by means of integration by parts, we get
$$
\C_\x(z|M)=-{\sin(F\pi)\over\pi^3}\,
{\cos(z\pi)\over1-2z}r^{2(z-1)}\int\limits_{|M|r}^\infty\,
{dw(w^2-M^2r^2)^{{1\over2}-z}\over \cosh[(2F-1)\ln({w\over\mu r})+\ln
A]}\times$$
$$\times\biggl\{
\bigl(F-{1\over2}\bigr)\bigl[K_F^2(w)-K_{1-F}^2(w)\bigr]
+2wK_F(w)K_{1-F}(w)+$$
$$+\bigl(F-{1\over2}\bigr)\bigl[K_F^2(w)+K_{1-F}^2(w)\bigr]
\tanh\bigl[(2F-1)\ln\bigl({w\over\mu r}\bigr)+\ln A\bigr]\biggr\},
\eqno(8.19)$$
where $\Re z<{3\over2}$ and
$$
\tilde{\zeta}_\x(z|M)={\sin(F\pi)\over\pi^3}\, {\sin(z\pi)\over1-z}
r^{2(z-1)}\int\limits_{|M|r}^\infty\, {dw\over w}(w^2-M^2r^2)^{1-z}\times$$
$$\times \biggl\{ {1\over2}\bigl[K_F^2(w)-K_{1-F}^2(w)\bigr]+
\bigl[FK_F^2(w)+(1-F)K_{1-F}^2+2w K_F(w)K_{1-F}(w)\bigr]\times$$
$$\times \tanh\bigl[(2F-1)\ln\bigl({w\over\mu r}\bigr)+\ln A\bigr]+
\bigl(F-{1\over2}\bigr)\bigl[K_F^2(w)+K_{1-F}^2(w)\bigr]
\tanh^2\bigl[(2F-1)\ln\bigl({w\over\mu r}\bigr)+\ln A\bigr]\biggl\},
\eqno(8.20)$$
where $\Re z<2$. Using the latter representations, we get
$$
\lim_{M\rightarrow 0}M\C_\x(z+{1\over2}|M)=\lim_{M\rightarrow
0}M^2\tilde{\zeta}_\x(z+1|M)=0, \quad \Re z<1, \eqno(8.21)$$
and, consequently,
$$
\lim_{M\rightarrow 0}\bpar\cdot\J^3(\x,\x;z|M)=2\tilde{\zeta}_\x(z|0),
\quad \Re z<1. \eqno(8.22)$$
One can easily get
$$
\tilde{\zeta}_\x(z|0)={\sin(F\pi)\over\pi^3}\sin(z\pi)r^{2(z-1)}\biggl\{
{\sqrt{\pi}\over2}\, {\Gamma(1-z)\over\Gamma({3\over2}-z)}
\bigl(F-{1\over2})\Gamma(F-z)\Gamma(1-F-z)+$$
$$+\int\limits_0^\infty dw\, w^{1-2z}\bigl[K_F^2(w)+K_{1-F}^2(w)\bigr]
\tanh\bigl[(2F-1)\ln\bigl({w\over\mu r}\bigr)+\ln A\bigr]\biggr\};
\eqno(8.23)$$
in particular, at half-integer values of the reduced vortex flux:
$$
\tilde{\zeta}_\x(z|0)\big|_{F={1\over2}}={s\sin\Theta\over
2\pi^{\sss3\over\sss2}}\, {\Gamma({1\over2}-z)\over\Gamma(z)} r^{2(z-1)};
\eqno(8.24)$$
and at $\cos\Theta=0$:
$$
\tilde{\zeta}_\x(z|0)=\pm{\sin(F\pi)\over2\pi^{\sss3\over\sss2}}\,
{\Gamma({3\over2}-z\pm F\mp{1\over2})\Gamma({1\over2}-z\mp
F\pm{1\over2})\over \Gamma(z)\Gamma({3\over2}-z)}\, r^{2(z-1)}, \quad
\Theta=\pm s{\pi\over2}. \eqno(8.25)$$
Consequently, we obtain
$$\tilde{\zeta}_\x(0|0)=0, \qquad \x\neq0. \eqno(8.26)$$

Thus the anomaly is absent everywhere on the plane with the puncture at
$\x=0$. This looks rather natural, since twodimensional anomaly
density $2\tilde{\zeta}_\x(0|0)$ is usually identified with
quantity ${s\over\pi}\bpar\times\V(\x)$ \cite{Schw,Shei,Jack}, and
the latter quantity in the present case vanishes everywhere on the
punctured plane, see Eq.(1.2). We see that the natural anticipations
are confirmed provided that the boundary conditions at the puncture
are chosen to be physically acceptable, i.e., compatible with the
self-adjointness of the Hamiltonian\footnote{The opposite claim of the
authors of Ref.\cite{Sold} is not justified.}; we conclude that
leakage of the anomaly, in contrast to that of the vacuum condensate 
or of the vacuum fermion number, does not happen.

We might finish here the discussion of the anomaly problem in the
background of a singular magnetic vortex. However, there remains a
purely academic question: what is the anomaly density in background 
(1.1) -- (1.2) on the whole plane (without puncturing
$\x=0$)? Just due to a confusion persisting in the literature
\cite{Sold,Mor}, we shall waste now some time to clarify this, otherwise
inessential, point.

Background field strength (1.2), when considered on the whole
plane, is interpreted in the sense of a distribution (generalized
function), i.e., a functional on a set of suitable test functions
$f(\x)$:
$$\int d^2x\,f(\x)\,{s\over\pi}\bpar\times\V(\x)=f(0) \,2 s\F ; \eqno(8.27)$$
here $f(\x)$ is a continuous function. In particular, choosing
$f(\x)=1$, one gets
$$\int d^2x\,{s\over\pi}\bpar\times\V(\x)=2 s\F. \eqno(8.28)$$
Considering the anomaly density on the whole plane, one is led to study
different limiting procedures as $r\rightarrow 0$ and $z\rightarrow 0$
in Eq.(8.23). So, the notorious question is, whether the anomaly
density $2\tilde{\zeta}_\x$ can be interpreted in the sense of a
distribution which coincides with distribution
${s\over\pi}\bpar\times\V(\x)$? The answer is resolutely negative, and
this will be immediately demonstrated below.

First, using explicit form (8.23), we get
$$
\i d^2x\,2\tilde{\zeta}_\x(z|0)=\left\{\ba{cc}
\infty,&z\neq0\\ 0,&z=0\\ \ea \right.\, ;\eqno(8.29)$$
therefore, the anomaly functional cannot be defined on the same set of
test functions as that used in Eq.(8.27) (for example, the test
functions have to decrease rapidly enough at large (small) distances in
the case of $z>0$ ($z<0$)). Moreover, if one neglects the requirement
of self-consistency, permitting a different (more specified) set of test 
functions for the anomaly functional, then even this will not save the 
situation. Let us take $z>0$ for definiteness and use the test functions 
which are adjusted in such a way that the quantity
$$
\A=\lim_{z\rightarrow 0_{+}}\i d^2x\,f(\x)\,2\tilde{\zeta}_\x(z|0)\eqno(8.30)$$
is finite. Certainly, this quantity can take values in a rather wide
range, but it cannot be made equal to the right-hand side of Eq.(8.28). 
Really, the only source of the dependence on $\F$ in the
integral in Eq.(8.30) is the factor $\tilde{\zeta}_\x(z|0)$, and the
latter, as is evident from Eq.(8.23), depends rather on $\fo\F-\Upsilon\fc$,
than on $\F$ itself, thus forbidding the linear dependence of $\A$ on
$\F$. In particular, let us choose test function $f(\x)$ in the form
$$
f(\x)=\exp(-{\tilde{\mu}}^{2}\,r^{2}), \eqno(8.31)$$
where $\tilde{\mu}$ is the parameter of the dimension of mass. Then, choosing 
the case of $\cos\Theta=0$ for simplicity and using Eq.(8.25), one gets that 
Eq.(8.30) takes the form
$$
\A=2\,\bigl[s(\fo\F-\Upsilon\fc-{1\over2})\pm{1\over2}\bigr],
\qquad \Theta=\pm s{\pi\over2}, \eqno(8.32)$$
which differs clearly from $2 s\F$.

Thus, in a singular background the conventional relation between the
anomaly density and the background field strength is valid only in
the space with punctured singularities. If the singularities are not
punctured, then the anomaly density and the background field strength
can be interpreted in the sense of distributions, but, in contrast to the
assertion of the authors of Refs.\cite{Sold,Mor}, the conventional
relation is not valid.

\section{Angular Momentum}

Let $\hat{M}$ be an operator in the first-quantized theory, which
commutes with the Dirac Hamiltonian
$$
[\hat{M},H]_-=0. \eqno(9.1)$$
Then, in the second-quantized theory, the vacuum expectation value of the
dynamical variable corresponding to $\hat{M}$ is presented in the form
$$
\M=\i d^2x\,\M_\x, \eqno(9.2)$$
where
$$
\M_\x=<\vac|{1\over2}\bigl[\Psi^+(\x,t),\hat{M}\,\Psi(\x,t)\bigr]_-|\vac>=
-{1\over2}\tr<\x|\hat{M}\,\sgn(H)|\x>. \eqno(9.3)$$
Commutation relation (9.1) is the evidence of invariance of the
theory with $\hat{M}$ being the generator of the symmetry
transformations. Since, in the background of a singular magnetic vortex
(1.1) -- (1.2), there is invariance with respect to  rotations
around the location of the vortex, one can take $\hat{M}$ as the
generator of rotations -- the operator of angular momentum in the
first-quantized theory (see \cite{SitR96} for more details):
$$\hat{M}=-i\x\times\bpar-\Upsilon+{1\over2}s\beta. \eqno(9.4)$$
Note that the eigenvalues of operator $\hat{M}$ (9.4) on spinor
functions satisfying condition (1.3) are half-integer.

Decomposing Eq.(9.4) into the orbital and spin parts, we get in the
second-quantized theory
$$
\M_\x=\L_\x+\S_\x, \eqno(9.5)$$
where
$$
\L_\x={1\over2}\tr<\x|(i\x\times\bpar+\Upsilon)\,\sgn(H)|\x> \eqno(9.6)$$
and
$$
\S_\x=-{1\over4}s\,\tr<\x|\beta\,\sgn(H)|\x>. \eqno(9.7)$$
Since vacuum spin density (9.7) is related to vacuum condensate (7.3),
$$
\S_\x={1\over2}s\,\C_\x, \eqno(9.8)$$
there remains only vacuum orbital angular momentum density (9.6) to
be considered.

Let us start, as in Section VII, with the regularized quantity
$$
\L_\x(z|M)={1\over2}\tr<\x|(i\x\times\bpar+\Upsilon)
H(H^2+M^2)^{-{1\over2}-z}|\x>. \eqno(9.9) $$
The contribution of regular solutions (2.17) and (2.18) to Eq.(9.9)
is cancelled upon summation over the sign of energy, whereas the
contribution of irregular solution (2.19) to Eq.(9.9) survives:
$$\L_\x(z|M)=-{1\over8\pi}\int\limits_0^\infty\, {dk\,
k^2(k^2+M^2)^{-{1\over 2}-z}}
\biggl\{ A\biggl({k\over\mu}\biggr)^{2F-1}\bigl[
\Lp+\Lm\bigr]\bigl[n_0J_{-F}^2(kr)+$$
$$+(n_0+s)J_{1-F}^2(kr)\bigr]+2\bigl[\Lp-\Lm\bigr]\bigl[n_0J_{-F}(kr)J_F(kr)
-(n_0+s)J_{1-F}(kr)J_{-1+F}(kr)\bigr]+$$
$$+A^{-1}\biggl({k\over\mu}
\biggr)^{1-2F} \bigl[\Lp+\Lm\bigr]\bigl[n_0J_F^2(kr)
+(n_0+s)J_{-1+F}^2(kr)\bigr]\biggr\}, \eqno(9.10)$$
where $A$ and $L_{(\pm)}$ are given by Eqs.(3.5) and (3.6).
Transforming the integral in Eq.(9.10) similarly to that as
in Appendix D, we get
$$\L_\x(z|M)=-{\sin(F\pi)\over2\pi^3}\cos(z\pi)
r^{2(z-1)}\int\limits_{|M|r}^\infty dw\,
w^2(w^2-M^2r^2)^{-{1\over2}-z}
 {n_0K_F^2(w)-(n_0+s)K_{1-F}^2(w)\over
\cosh[(2F-1)\ln({w\over\mu r})+\ln A]}, \eqno(9.11)$$
where $\Re z<{1\over2}$. Then we get
$$
\L_\x\equiv \L_\x(0|0)=-{\sin(F\pi)\over2\pi^3r^2}\int\limits_0^\infty dw\, w
{n_0K_F^2(w)-(n_0+s)K_{1-F}^2(w)\over \cosh[(2F-1)\ln({w\over\mu
r})+\ln A]}. \eqno(9.12)$$

Summing Eqs.(9.12) and (9.8), taking into account Eqs.(7.7) and
(2.13), we obtain the following expression  for the vacuum angular
momentum density in the background of a singular magnetic vortex
(1.1) -- (1.2):
$$\M_\x=\bigl(\io\F-\Upsilon\ic+{1\over2}\bigr)\,\N_\x, \eqno(9.13)$$
where vacuum fermion number density $\N_\x$ is given by Eq.(3.7).
Thus, the total vacuum angular momentum takes the form (see Eq.(3.9))
$$\M=-{1\over2}\bigl(\io\F-\Upsilon\ic+{1\over2}\bigr)\,
\sgn_0\bigl[(F-{1\over2})\cos\Theta\bigr]. \eqno(9.14)$$

Concluding this section, let us note that relation (9.1) remains to
be valid if a constant is added to operator $\hat{M}$. Thus, a
definition which is alternative to Eq.(9.4) has been proposed for the
angular momentum in the first-quantized theory \cite{Wil}:
$$\hat{M}'=-i\x\times\bpar-\F+{1\over2}s\beta. \eqno(9.15)$$
Then in the second-quantized theory we get
$$\M'_\x=-s\,\bigl(F-{1\over2}\bigr)\,\N_\x \eqno(9.16)$$
and
$$\M'={1\over2}\,s\,|F-{1\over2}|\,\sgn_0(\cos\Theta).\eqno(9.17)$$
Various arguments pro and contra the physical meaningfulness of 
operator $\hat{M}'$ (9.15) are known in the literature (see Refs.
\cite{Wil,Jack1,Jack2}). However, the crucial point is that 
operator $\hat{M}$ is the generator of rotations, while operator 
$\hat{M}'$ is not (the eigenvalues of operator $\hat{M}'$ on spinor 
functions are not half-integer).

\section{Summary and discussion}

In the present paper we show that the massless fermionic vacuum under
the influence of a singular magnetic vortex (1.1) -- (1.2) in 
2+1-dimensional space-time attains the following nontrivial
characteristics: fermion number (density (3.7) and total (3.9)), energy
density (5.2), current (6.14), parity breaking condensate (density
(7.7) and total (7.10)) and angular momentum (density (9.13) and total
(9.14)). The vacuum spin is related to the vacuum condensate, see
Eq.(9.8), the effective potential coincides with the vacuum energy
density, see Eq.(5.19), and the parity  anomaly is absent, see
Eq.(8.8). At large distances from the vortex local vacuum
characteristics (densities) are decreasing as inverse powers (with
integer exponents in the cases of energy density and current, and with
fractional exponents otherwise), see Eqs.(3.8), (5.5), (6.18), and
(7.9).

The most general set of boundary conditions at the location of the vortex 
is used, see Eq.(2.14), providing the self-adjointness of the Dirac
Hamiltonian;  thus all vacuum polarization effects are depending on
self-adjoint extension parameter $\Theta$ or $A$ (3.5). As to the
dependence on vortex flux $\F$, it has been already anticipated in 
Introduction that all vacuum polarization effects are gauge invariant 
and thus depend on 
reduced vortex flux $\F-\Upsilon$ rather than on $\F$
or $\Upsilon$ separately. To be more precise, all effects are 
certainly depending on the fractional part of $\F-\Upsilon$ 
(i.e. on $F$ (2.15)). It is clear that the dependence on
the integer part of $\F-\Upsilon$ can be achieved with the help of
boundary condition, i.e. choosing the value of $\Theta$ differently for
different values of $\io\F-\Upsilon\ic$. However, the vacuum angular
momentum, in contrast to all other vacuum characteristics, yields a
pattern of other, much more essential, dependence on
$\io\F-\Upsilon\ic$. This is owing to the remarkable linear relation
between the vacuum angular momentum and fermion number, see Eq.(9.13),
which in its turn is due to the choice of the generator of rotations in
the capacity of the operator of angular momentum in the
first-quantized theory, see Eq.(9.4). Although all vacuum
characteristics are vanishing at integer values of $\F-\Upsilon$ (i.e.
at $F={1\over2}-{1\over2}s$ ),\footnote{This confirms once more the
general fact that a singular magnetic vortex is physically unobservable
at integer values of $\F-\Upsilon$ \cite{Aha}. It was as far back as
1931 that Dirac used actually this fact to obtain his famous condition
for the magnetic monopole quantization \cite{Dir}.} at noninteger and
non-half-integer values (i.e. at $0<F<{1\over2}$ and at
${1\over2}<F<1$) the absolute  value of the vacuum angular momentum, in
contrast to all other vacuum characteristics, is increasing linearly
with the increase of the absolute value of $\io\F-\Upsilon\ic$, see
Eqs.(9.13) and (9.14). Thus, if the vacuum angular momentum could
become somehow detectable, then this would provide us with a unique
explicit evidence in favour of physical effects that depend
essentially both on integer and fractional parts of the enclosed
magnetic flux. Note also that at half-integer values of $\F-\Upsilon$
(i.e. at $F={1\over2}$) the vacuum fermion number and angular momentum
are vanishing, whereas the vacuum energy density, current and
condensate are nonvanishing (the total condensate is even infinite),
see Eqs.(5.3), (6.17), (7.8) and (7.10), unless one chooses a certain
boundary condition to be discussed immediately below.

Among the whole variety of boundary conditions which are specified by
self-adjoint extension parameter $\Theta$, condition
$\cos\Theta=0$ (or $\Theta=\pm{\pi\over2}$) is distinguished, since it
corresponds to one of the two components of a solution to the Dirac
equation being regular for all $n$: if $\Theta=s{\pi\over2}$, then the
lower components are regular, and, if $\Theta=-s{\pi\over2}$, then the
upper components are regular, see Eqs.(2.17) -- (2.20). This condition 
is parity invariant, and under it
the vacuum fermion number, condensate and angular momentum
are vanishing, whereas the vacuum energy density and current are
nonvanishing, see Eqs.(5.4) and (6.15). It should be noted that this
condition is extensively discussed in the literature, being involved into
the two most popular ones: the condition of maximal simplicity
\cite{Alf}

$$\Theta=\left\{\ba{cc}
s{\pi\over2},& s(\F-\Upsilon)>0\\[0.2cm]
-s{\pi\over2},& s(\F-\Upsilon)<0\\ \ea \right\} \eqno(10.1)
$$
and the condition of minimal irregularity \cite{Sit90,Sit96}

$$\Theta=\left\{\ba{cc}
s{\pi\over2},& -{1\over2}<s(\fo\F-\Upsilon\fc-{1\over2})<0\\[0.2cm]
0,& \fo\F-\Upsilon\fc={1\over2}\\[0.2cm]
-s{\pi\over2},& 0<s(\fo\F-\Upsilon\fc-{1\over2})<{1\over2}\\ \ea
\right\}; \eqno(10.2)
$$
here, both in Eqs.(10.1) and(10.2), it is implied 
that $\fo\F-\Upsilon\fc\neq0$.

Under condition (10.1) we get

$$
\E_\x^\ren=\left\{\ba{cc}
{\tan(\fo\F-\Upsilon\fc\pi)\over12\pi r^3}
\fo\F-\Upsilon\fc\bigl({1\over4}-\fo\F-\Upsilon\fc^2\bigr),&
\F-\Upsilon>0\\[0.2cm]
{\tan[(1-\fo\F-\Upsilon\fc)\pi]\over 12\pi
r^3}\bigl(1-\fo\F-\Upsilon\fc\bigr)
\bigl[{1\over4}-(1-\fo\F-\Upsilon\fc)^2\bigr],& \F-\Upsilon<0\\ \ea
\right\}
\eqno(10.3) $$
and
$$
j_\varphi(\x)=\left\{\ba{cc}
-{\tan(\fo\F-\Upsilon\fc\pi)\over4\pi
r^2}\bigl({1\over4}-\fo\F-\Upsilon\fc^2\bigr),& \F-\Upsilon>0\\[0.2cm]
{\tan[(1-\fo\F-\Upsilon\fc)\pi]\over4\pi r^2}
\bigl[{1\over4}-(1-\fo\F-\Upsilon\fc)^2\bigr],& \F-\Upsilon<0\\ \ea
\right\}. \eqno(10.4)
$$
Under condition (10.2) we get
$$
\E_\x^\ren=\left\{ \ba{cc}
{\tan(\fo\F-\Upsilon\fc\pi)\over24\pi r^3}
\bigl(\fo\F-\Upsilon\fc-{1\over2}\bigr)
\bigl[3|\fo\F-\Upsilon\fc-{1\over2}|-\\[0.2cm]
-2(\fo\F-\Upsilon\fc-{1\over2})^2-1\bigr],
& \fo\F-\Upsilon\fc\neq{1\over2}\\[0.2cm]
{1\over24\pi^2 r^3},& \fo\F-\Upsilon\fc={1\over2}\\ \ea
\right\}, \eqno(10.5)
$$
$$
j_\varphi(\x)=\left\{\ba{cc}
{\tan(\fo\F-\Upsilon\fc\pi)\over4\pi r^2}\big|\fo\F-\Upsilon\fc-{1\over2}\big|
\bigl(\big|\fo\F-\Upsilon\fc-{1\over2}\big|-1\bigr),&
\fo\F-\Upsilon\fc\neq{1\over2}\\[0.2cm]
0,& \fo\F-\Upsilon\fc={1\over2}\\ \ea \right\}, \eqno(10.6)
$$
$$
\C_\x=\left\{ \ba{cc}
0,& \fo\F-\Upsilon\fc\neq{1\over2}\\[0.2cm]
-{1\over 4\pi^2 r^2},& \fo\F-\Upsilon\fc={1\over2}\\ \ea
\right\} \eqno(10.7)
$$
and
$$
\C=\left\{ \ba{cc}
0,&\fo\F-\Upsilon\fc\neq{1\over2}\\[0.2cm]
-\infty,& \fo\F-\Upsilon\fc={1\over2}\\ \ea \right\}. \eqno(10.8)
$$
It is clear that Eqs.(10.5) -- (10.8), in contrast to Eqs.(10.3) --
(10.4), are periodic in the value of the vortex flux, however, 
they, excepting $\E_\x^\ren$ (10.5), have discontinuities 
(even an infinite jump in $\C$ (10.8)) at both sides
of half-integer values of the reduced vortex flux
($\fo\F-\Upsilon\fc={1\over2}$).

As it should be expected, all vacuum polarization effects are
invariant under transitions to equivalent representations of the
Clifford algebra (i.e. independent of $\chi_s$). Under the transition
to an inequivalent representation (i.e. under $s\rightarrow -s$) the
vacuum fermion number and angular momentum change their sign, while the
vacuum energy density, current and condensate remain unchanged. The
latter fact has immediate but far reaching consequences for the more
realistic 2+1-dimensional fermionic model without parity violation.

In this model a quantized spinor field is assigned to a reducible
representation composed of two inequivalent irreducible  ones. Namely,
the Dirac $\gamma$ matrices are chosen in the form

$$
\Gamma^0=\left(\ba{cc}
\gamma^0&0\\
0&-\gamma^0\\ \ea\right), \qquad \bGamma=\left(\ba{cc}
\gb& 0\\
0&-\gb\\ \ea \right), \eqno(10.9)
$$
where $\gamma^0$ and $\gb$ are the $2\times2$ matrices of the
irreducible representation, see Eqs.(2.7) -- (2.9), and a
four-component spinor $\Xi$ is a doublet of two-component spinors
$\Psi_+$ and $\Psi_-$:

$$
\Xi=\left(\ba{c} \Psi_+\\ \Psi_-\\ \ea\right). \eqno(10.10)
$$
Term $\Xi^+\Gamma^0\Xi$ is invariant under all types of
parity transformations (including space and time reflections) but
violates chiral symmetry (see, for example, Ref. \cite{Appe}). Such
four-component quantized spinor fields may be relevant for planar
condensed matter systems with two sublattices, exhibiting type-II
superconductivity \cite{Dor}. Our concern is in the following: how the
ground state is affected by an external field configuration in the form
of a singular magnetic vortex (1.1) -- (1.2)?

Actually, the vacuum polarization effects which are due to $\Psi_+$
(upper component of doublet (10.10)) have been already determined
in the present paper. To determine the vacuum polarization effects
which are due to $\Psi_-$ (lower component of doublet (10.10)), one
has to repeat all calculations after changing $\gamma^0\rightarrow
-\gamma^0$, $\gamma^1\rightarrow -\gamma^1$, $\gamma^2\rightarrow
-\gamma^2$, see Eq.(10.9). Doing this, we find that the latter effects
are equal to the former ones after changing $s\rightarrow -s$. Thus,
polarization of the vacuum for $\Xi$ (10.10) is obtained just by
summing polarization of the vacuum for $\Psi$ (2.1) over $s=\pm1$. In
particular, generalized current $\sum_{s=\pm1}{\bf J}^3$, in
contrast  to $\sum_{s=\pm1}{\bf J}$, vanishes identically (see
Eq.(8.2)), and the anomaly problem which was worrying us along the
whole Section VIII is absent at all. Also in this case the vacuum spin, as
well as the vacuum angular momentum and fermion number, vanishes
identically (see Eq.(9.8)), whereas the vacuum condensate survives but
exhibits chiral symmetry breaking, since it corresponds to the vacuum
expectation value of $\Xi^+\Gamma^0\Xi$. Summarizing, the
nontrivial vacuum characteristics are:

energy density
$$
\sum_{s=\pm1}\,\E_\x^\ren={\sin(\fo\F-\Upsilon\fc\pi)\over\pi r^3}\,
\biggl\{ { {1\over2}-\fo\F-\Upsilon\fc\over
6\cos(\fo\F-\Upsilon\fc\pi)} \left[{3\over4}
-\fo\F-\Upsilon\fc(1-\fo\F-\Upsilon\fc)\right]+$$
$$+{1\over\pi^2}\int\limits_0^\infty dw
w^2\bigl[K^2_{\fo\F-\Upsilon\fc}(w)-K^2_{1-\fo\F-\Upsilon\fc}(w)\bigr]
\tanh\bigl[(2\fo\F-\Upsilon\fc-1)\ln({w\over\mu
r})+\ln\bar{A}\bigr]\biggr\}, \eqno(10.11)
$$

current
$$
\sum_{s=\pm1}j_\varphi(\x)={2\sin(\fo\F-\Upsilon\fc\pi)\over\pi r^2}
\biggl\{ {(\fo\F-\Upsilon\fc-{1\over2})^2\over
4\cos(\fo\F-\Upsilon\fc\pi)}-$$
$$-{1\over\pi^2}\int\limits_0^\infty dw w K_{\fo\F-\Upsilon\fc}(w)
K_{1-\fo\F-\Upsilon\fc}(w)\tanh \bigl[(2\fo\F-\Upsilon\fc-1)
\ln({w\over\mu r})+\ln\bar{A}\bigr]\biggr\}, \eqno(10.12)
$$

condensate density
$$
\sum_{s=\pm1}\,\C_\x=-{\sin(\fo\F-\Upsilon\fc\pi)\over\pi^3 r^2}
\int\limits_0^\infty dw w
{K^2_{\fo\F-\Upsilon\fc}(w)+K^2_{1-\fo\F-\Upsilon\fc}(w)\over
\cosh[(2\fo\F-\Upsilon\fc-1)\ln({w\over\mu r})+\ln\bar{A}]}
\eqno(10.13)
$$

and condensate total
$$
\sum_{s=\pm1}\C=-{\sgn_0(\cos\Theta)\over2|\fo\F-\Upsilon\fc-{1\over2}|},
\eqno(10.14)
$$
where
$$
\bar{A}=2^{1-2\fo\F-\Upsilon\fc}\,
{\Gamma(1-\fo\F-\Upsilon\fc)\over\Gamma(\fo\F-\Upsilon\fc)}
\tan\bigl({\Theta\over2}+{\pi\over4}\bigr). \eqno(10.15)
$$                                               
It should be
noted that chiral symmetry breaking in the background of regular
configurations of an external magnetic field has been extensively
discussed in the literature \cite{Gus,Dunn}. We conclude that chiral
symmetry breaking occurs also in the background of a singular
configuration of an external magnetic field, as a result of leakage
of the condensate from the singularity point.

\section*{Acknowledgements}

I thank H.Leutwyler for interesting discussions. This 
research was supported by the State Foundation for Fundamental
Research of Ukraine (project 2.4/320) and the Swiss National Science
Foundation (grant CEEC/NIS/96-98/7 IP 051219).

\section*{Appendix A}
\def\theequation{A.\arabic{equation}}
\setcounter{equation}{0}

In view of the discussion in Section IV, we consider here a more
general case of massive Hamiltonian $\tilde{H}$ (4.6). The relevant
partial Hamiltonian has the form
\be
\tilde{h}_{n_0}=\left(\begin{array}{cc}m&e^{i\chi_s}[\partial_r+(1-F)r^{-1}]\\
e^{-i\chi_s}(-\partial_r-Fr^{-1})&-m\end{array}\right).
\ee
Let $\tilde{h}$ be the operator in the form of Eq.(A.1), which acts on
the domain of functions
$\xi^0(r)$ that are regular at $r=0$. Then its adjoint $\tilde{h}^{\dagger}$
which is defined by the relation
\be
\int^\infty_0 dr\,r[\tilde{h}^\dagger\xi(r)]^\dagger \xi^0(r) =
\int^\infty_0 dr\,r[\xi(r)]^\dagger[\tilde{h}\xi^0(r)]
\ee
acts on the domain of functions $\xi(r)$ that are not necessarily
regular at $r=0$. So the question is, whether the domain of definition
of $\tilde{h}$ can be extended, resulting in both the operator and its
adjoint being defined on the same domain of functions? To answer this,
one has to construct the eigenspaces of $\tilde{h}^\dagger$
with complex eigenvalues. They are spanned by the linearly independent
square-integrable solutions correspoding to the pair of purely imaginary
eigenvalues,
\be
\tilde{h}^\dagger \xi^\pm (r) = \pm i\mu\xi^\pm (r),
\ee
where $\mu>0$ is inserted for the dimension reasons. It can be shown that in the
case of Eq.(A.1) only one pair of such solutions exists, thus the deficiency
index of $\tilde{h}$ is equal to (1,1). This pair is given by the following
expression
\be
\xi^\pm (r) = {1\over N}\left(\begin{array}{l}e^{i\chi_s} \exp{\bigl[\pm{i\over2}
\sgn (m) \eta\bigr]}K_F(\tilde{\mu} r)\\
\sgn (m) \exp{\bigl[\mp{i\over2}\sgn (m) \eta\bigr]}K_{1-F}(\tilde{\mu}
r)\end{array}\right),
\ee
where $N$ is a certain normalization factor and
\be
\tilde{\mu} = \sqrt{\mu^2 + m^2},\,\,  \eta = \arctan\biggl({\mu\over|m|}\biggr).
\ee
Self-adjoint extended operator $\tilde{h}^{\theta_s}$ is defined on
the domain of functions of the form
\be
\left(\ba{c}
\tilde{f}_{n_0}\\
\tilde{g}_{n_0}\\ \ea
\right) = \xi^0 + c(\xi^+ + e^{i\theta_s}\xi^-),
\ee
where $c$ is a complex parameter and $\theta_s$ is a real continuous
parameter which depends, in general, on the choice between the two
inequivalent representations of the Clifford algebra. Using the
asymptotics of the Macdonald function at small values of the variable,
we get
\be
\left(\begin{array}{c}\tilde{f}_{n_0}\\ \tilde{g}_{n_0}\end{array}\right)
\mathop{\sim}_{r\to0}
\left(\begin{array}{l}e^{i\chi_s} \cos\left\{{1\over2}[\theta_s-\sgn (m)
\eta]\right\}
2^F\Gamma (F)(\tilde{\mu} r)^{-F}\\
\sgn (m)\cos\left\{{1\over2}[\theta_s+\sgn (m)\eta]\right\}2^{1-F}\Gamma
(1-F)(\tilde{\mu}r)^{-1+F}\end{array}\right),
\ee
or
\bea
\left\{\tan\big[{1\over2}\theta_s - {1\over2}\sgn (m) \eta\big] \sin \eta
- \sgn (m)\cos\eta\right\} \lim_{r\to 0}(\tilde{\mu}r)^F\tilde{f}_{n_0}=
\nonumber
\eea
\bea=
-e^{i\chi_s}2^{2F-1}{\Gamma(F)\over\Gamma(1-F)}\lim_{r\to 0}(\tilde\mu
r)^{1-F}\tilde{g}_{n_0}.
\eea
Defining new parameter $\Theta$ by means of relation
\bea
&& \tan\biggl(s{\Theta\over2} +{\pi\over4}\biggr)=\left\{\tan\big[{1\over2}
\theta_s-{1\over2}\sgn(m)\eta\big]\sin\eta-\sgn(m)\cos\eta\right\}^{-1}
 \biggl({2\mu\over\tilde{\mu}}\biggr)^{2F-1}{\Gamma(F)\over\Gamma(1-F)},
\eea
we get Eq.(2.14).

Certainly, both $\theta_s$ and $\Theta$ can be regarded as self-adjoint
extension parameters which are to specify the boundary condition at $r=0$. The
use of $\Theta$ in this aspect may seem to be more preferable just for
the convenience reasons, because Eq.(2.14) looks much simpler than
Eq.(A.8). In particular, Eq.(2.14), in contrast to Eq.(A.8), is independent
of $m$ and remains explicitly invariant under $s\to-s$ ($\Theta$ is
independent of $s$).

In the limit of $m\to \pm 0$ Eq.(A.9) takes the form
\be
\tan\biggl(s{\Theta\over2}+{\pi\over4}\biggr)=-\tan\big[{1\over2}
\theta_s+{1\over4}\sgn(m)\pi\big]2^{2F-1}{\Gamma(F)\over\Gamma(1-F)},
\ee
then we get
\be
\theta_s=s\theta,
\ee
where $\theta$ is independent of s.

Concluding this appendix, let us note that all characteristics of the
massless fermionic vacuum are depending on $A$ (3.5) rather than on
$\Theta$ itself, and $A$ is expressed through $\theta$ in the following way
\be
A=-\tan\big[{1\over2}s\theta + {1\over4}\sgn(m) \pi\big].
\ee

                                                                   %\cl
\section*{Appendix B}
\def\theequation{B.\arabic{equation}}
\setcounter{equation}{0}

With the help of relations (see, for example, Ref. \cite{Abra})

$$J_\rho(iz)=\exp\bigl({i\over2}\rho\pi\bigr)I_\rho(z), \qquad
-\pi<\arg\,z\leq{\pi\over2},
$$
$$I_\rho(-z)=\exp\bigl(i\rho\pi\bigr)I_\rho(z), \quad
K_\rho(-z)=\exp\bigl(-i\rho\pi\bigr)K_\rho(z)-i\pi I_\rho(z), \quad
-\pi<\arg\,z<0,
$$
we get
\bea
&&J_\rho(kr)J_\tau(kr)={1\over2i\pi}\bigl\{
\exp\bigl[{i\over2}(\rho-\tau)\pi\bigr]I_\rho(-ikr)K_\tau(-ikr)-\nonumber \\
&&-\exp\bigl[{i\over2}(\tau-\rho)\pi\bigr]I_\rho(ikr)K_\tau(ikr)+
\exp\bigl[
{i\over2}(\tau-\rho)\pi\bigr]I_\tau(-ikr)K_\rho(-ikr)-\nonumber \\
&&-\exp\bigl[{i\over2}(\rho-\tau)\pi\bigr]I_\tau(ikr)K_\rho(ikr)\bigr\}.
\label{B.1}
\eea
Then, using Eq.(B.1), we recast Eq.(3.4) into the form
\be
\N_\x=\int\limits_{C_{\mbox{\DamirFont\symbol{'001}}}}
d\omega{\cal F}_1(\omega).
\label{B.2}
\ee
Here $\omega=k^2$ is the new variable of integration; contour
${C_{\mbox{\DamirFont\symbol{'001}}}}$
circumvents the real positive semiaxis of variable $\omega$, going
along it at infinitely small distances from below and above; and the
integrand has the form

\bea
&&\Fi_1(\omega)={i\over(4\pi)^2} \bigl\{
A\bigl({\omega\over\mu^2}\bigr)^{F-{1\over2}}[\Lp+\Lm] 
[I_{-F}(r\sqrt{-\omega})K_F(r\sqrt{-\omega})+\nonumber\\
&&+I_{1-F}(r\sqrt{-\omega})K_{1-F}(r\sqrt{-\omega})]+\omega^F [\Lp-\Lm]
(-\omega)^{-F}I_F(r\sqrt{-\omega})K_F(r\sqrt{-\omega})+\nonumber\\
&&+\omega^{-F}[\Lp-\Lm](-\omega)^F
I_{-F}(r\sqrt{-\omega})K_F(r\sqrt{-\omega})-\nonumber \\
&&-\omega^{1-F}
[\Lp-\Lm](-\omega)^{-1+F} I_{1-F}(r\sqrt{-\omega})K_{1-F}
(r\sqrt{-\omega})-\nonumber \\
&&-\omega^{-1+F} [\Lp-\Lm] (-\omega)^{1-F}
I_{-1+F}(r\sqrt{-\omega})K_{1-F}(r\sqrt{-\omega})+\nonumber \\
&&+A^{-1}\bigl({\omega\over\mu^2}\bigr)^{{1\over2}-F} [\Lp+\Lm]
[I_F(r\sqrt{-\omega})K_F(r\sqrt{-\omega})+I_{-1+F}(r\sqrt{-\omega})
K_{1-F}(r\sqrt{-\omega})\bigr]\bigr\}.
\label{B.3}
\eea
By continuously deforming the contour of integration in the complex
$\omega$-plane as is shown in Fig.1, we arrive at the relation

\be
\int\limits_{C_{\mbox{\DamirFont\symbol{'001}}}}d\omega{\cal
F}_1(\omega)=
\int\limits_{C_{\mbox{\DamirFont\symbol{'005}}}}d\omega{\cal
F}_1(\omega)+
\int\limits_{C_{\mbox{\DamirFont\symbol{'002}}}}d\omega{\cal
F}_1(\omega)+
\int\limits_{C_{\mbox{\DamirFont\symbol{'004}}}}d\omega{\cal
F}_1(\omega). \label{B.4} \ee
The integrals along semicircles
$C_{\mbox{\DamirFont\symbol{'005}}}$ and
$C_{\mbox{\DamirFont\symbol{'004}}}$ of infinite radii vanish, whereas
the integral along the contour circumventing the real negative semiaxis
can be represented as

\bea
&&\int\limits_{C_{\mbox{\DamirFont\symbol{'002}}}}d\omega{\cal F}_1(\omega)
=-{1\over(4\pi)^2} \int\limits_0^\infty du\times\nonumber \\
&&\times\biggl(
A\bigl({u\over\mu^2}\bigr)^{F-{1\over2}}
\bigl\{e^{iF\pi}\bigl[\Rp+\Rpm\bigr]+ e^{-iF\pi}
\bigl[\Rmp+\Rm\bigr]\bigr\}\times\nonumber\\
&&\times
\bigl[I_{-F}(r\sqrt{u})K_F(r\sqrt{u})+I_{1-F}(r\sqrt{u})K_{1-F}(r\sqrt{u})
\bigr]-\nonumber \\
&&-\bigl\{ e^{i(F-{1\over2})\pi} \bigl[\Rp-\Rpm\bigr]- 
e^{-i(F-{1\over2})\pi} \bigl[\Rmp-\Rm\bigr]\bigr\}\times \nonumber\\
&&\times \bigl[I_F(r\sqrt{u})K_F(r\sqrt{u})+I_{-1+F}(r\sqrt{u})
K_{1-F}(r\sqrt{u})\bigr]+\nonumber \\
&&+\bigl\{e^{-i(F-{1\over2})\pi}\bigl[\Rp-\Rpm\bigr]
-e^{i(F-{1\over2})\pi}\bigl[\Rmp-\Rm\bigr]\bigr\} \times \nonumber\\
&&\times \bigl[I_{-F}(r\sqrt{u})K_F(r\sqrt{u})+ 
I_{1-F}(r\sqrt{u})K_{1-F}(r\sqrt{u})\bigr]-\nonumber \\
&&-A^{-1}\bigl({u\over\mu^2}\bigr)^{{1\over2}-F} \bigl\{
e^{-iF\pi}\bigl[\Rp+\Rpm\bigr]+ e^{iF\pi}\bigl[\Rmp+\Rm\bigr]\bigr\}\times\nonumber\\
&&\times \bigl[I_F(r\sqrt{u})K_F(r\sqrt{u})+I_{-1+F}(r\sqrt{u})
K_{1-F}(r\sqrt{u})\bigr]\biggr),
\label{B.5}
\eea
where

\be
R_{(\pm)}^{(+)}=2^{-1}\biggl(\cos(F\pi)\pm \cosh\bigl\{\bigl(F-{1\over2}
\bigr)\bigl[ \ln\bigl({u\over\mu^2}\bigr)+i\pi\bigr]+\ln\,
A\bigr\}\biggr)^{-1}
\label{B.6}
\ee
and
\be
R_{(\pm)}^{(-)}=2^{-1}\biggl(\cos(F\pi)\mp \cosh\bigl\{
\bigl(F-{1\over2}\bigr)\bigl[\ln\bigl({u\over\mu^2}\bigr)-i\pi\bigr]+
\ln\, A\bigr\} \biggr)^{-1}.
\label{B.7}
\ee
Expression (B.5) can be reduced to the form

\bea
&&\int\limits_{C_{\mbox{\DamirFont\symbol{'002}}}}d\omega{\cal F}_1(\omega)
=-{\sin(F\pi)\over 8\pi^3}\int\limits_0^\infty du\bigl\{
A\bigl({u\over\mu^2}\bigr)^{F-{1\over2}}
\bigl[e^{iF\pi}\bigl(\Rp+\Rpm\bigr)
+e^{-iF\pi}\bigl(\Rmp+\Rm\bigr)\bigr]+\nonumber \\
&&+ e^{-i(F-{1\over2})\pi}\bigl(\Rp-\Rpm\bigr) -e^{i(F-{1\over2})\pi}
\bigl(\Rmp-\Rm\bigr)\bigr\}K_F^2(r\sqrt{u})+\nonumber \\
&& +{\sin(F\pi)\over 8\pi^3} \int\limits_0^\infty du\bigl\{
e^{i(F-{1\over2})\pi}\bigl(\Rp-\Rpm\bigr) -e^{-i(F-{1\over2})\pi}
\bigl(\Rmp-\Rm\bigr)+\nonumber \\
&&+A^{-1}\bigl({u\over\mu^2}\bigr)^{{1\over2}-F}
\bigl[e^{-iF\pi}\bigl(\Rp+\Rpm\bigr)+e^{iF\pi}\bigl(\Rmp+\Rm\bigr)\bigr]
\bigr\}K_{1-F}^2(r\sqrt{u}).
\label{B.8}
\eea
After further simplifications we arrive at Eq.(3.7) where variable
$w=r\sqrt{u}$ is introduced.

%\cl
\section*{Appendix C}
\def\theequation{C.\arabic{equation}}
\setcounter{equation}{0}

As in the previous appendix, with the use of Eq.(B.1), expressions
(4.17) and (4.18) are recast into the form

\be
\bigl[\zeta_\x(z|m)\bigr]_\REG=
\int\limits_{C_{\mbox{\DamirFont\symbol{'001}}}}d\omega{\cal
F}_2(\omega) \label{C.1}
\ee
and
\be \bigl[\zeta_\x(z|m)\bigr]_\IRREG=
\int\limits_{C_{\mbox{\DamirFont\symbol{'001}}}}d\omega{\cal
F}_3(\omega),
\label{C.2}
\ee
where

\be
{\cal F}_2(\omega)={1\over i(2\pi)^2}\,
{(\omega+m^2)^{1-z}\over(z-1)\omega}\bigl[
FI_F(r\sqrt{-\omega})K_F(r\sqrt{-\omega})+
(1-F)I_{1-F}(r\sqrt{-\omega}) K_{1-F}(r\sqrt{-\omega})\bigr]
\label{C.3}
\ee
and
\bea
&& {\cal F}_3(\omega)= {1\over i2(2\pi)^2}\,
(\omega+m^2)^{-{1\over2}-z} \bigl\{
A\mu^{1-2F}\omega^F\bigl[\tLp-\tLm\bigr]
I_{-F}(r\sqrt{-\omega})K_F(r\sqrt{-\omega})+\nonumber \\
&&+ A\mu^{1-2F}\omega^{-1+F}\bigl[ (m-\sqrt{\omega+m^2})^2\tLp
-(m+\sqrt{\omega+m^2})^2 \tLm\bigr]
I_{1-F}(r\sqrt{-\omega})K_{1-F}(r\sqrt{-\omega})+\nonumber \\
&&+
\omega^F\bigl[(m+\sqrt{\omega+m^2})\tLp-(m-\sqrt{\omega+m^2})\tLm\bigr]
(-\omega)^{-F} I_F(r\sqrt{-\omega})K_F(r\sqrt{-\omega})+\nonumber \\
&&+\omega^{-F}\bigl[(m+\sqrt{\omega+m^2})\tLp-(m-\sqrt{\omega+m^2})\tLm
\bigr] (-\omega)^F
I_{-F}(r\sqrt{-\omega})K_F(r\sqrt{-\omega})+\nonumber \\
&&+ \omega^{1-F}\bigl[
(m-\sqrt{\omega+m^2})\tLp-(m+\sqrt{\omega+m^2})\tLm\bigr]
(-\omega)^{-1+F}
I_{1-F}(r\sqrt{-\omega})K_{1-F}(r\sqrt{-\omega})+\nonumber \\
&&+\omega^{-1+F}\bigl[
(m-\sqrt{\omega+m^2})\tLp-(m+\sqrt{\omega+m^2})\tLm\bigr]
(-\omega)^{1-F}
I_{-1+F}(r\sqrt{-\omega})K_{1-F}(r\sqrt{-\omega})+\nonumber \\
&&+A^{-1}\mu^{2F-1}\omega^{-F}\bigl[ (m+\sqrt{\omega+m^2})^2\tLp-
(m-\sqrt{\omega+m^2})^2\tLm\bigr]
I_F(r\sqrt{-\omega})K_F(r\sqrt{-\omega})+\nonumber \\
&&+ A^{-1}\mu^{2F-1}\omega^{1-F}\bigl[ \tLp-\tLm\bigr]
I_{-1+F}(r\sqrt{-\omega})K_{1-F}(r\sqrt{-\omega})\bigr\}.
\label{C.4}
\eea
Deforming the contour of integration in Eqs.(C.1) and (C.2) as is
shown in Figs. 2 and 3 correspondingly, we arrive at the relation

\be
\int\limits_{C_{\mbox{\DamirFont\symbol{'001}}}}d\omega{\cal
F}_j(\omega) =
\int\limits_{C_{\mbox{\DamirFont\symbol{'005}}}}d\omega{\cal
F}_j(\omega) +
\int\limits_{C_{\mbox{\DamirFont\symbol{'002}}}}d\omega{\cal
F}_j(\omega) +
\int\limits_{C_{\mbox{\DamirFont\symbol{'004}}}}d\omega{\cal
F}_j(\omega) +
\int\limits_{C_{\mbox{\DamirFont\symbol{'011}}}}d\omega{\cal
F}_j(\omega),
\quad j=2,3.
\label{C.5}
\ee
Here both integrands (C.3) and (C.4) have a branching point at
$\omega=-m^2$ (therefore a cut is drawn from $-m^2$ to $-\infty$) and a
pole on the real axis: at $\omega=0$ in the case of Eq.(C.3), and at
$\omega=-\kappa^2$ in the case of Eq.(C.4), where $\kappa^2=m^2-E_{\rm
BS}^2$ and $E_{\rm BS}$ is determined by Eq.(4.14). At $1<\Re z<2$ 
in the case of ${\cal F}_2(\omega)$ and at ${1\over2}<\Re z<1$ in
the case of ${\cal F}_3(\omega)$, the integrals along semicircles
$C_{\mbox{\DamirFont\symbol{'005}}}$ and
$C_{\mbox{\DamirFont\symbol{'004}}}$ of infinite radii vanish, whereas
the integrals along the contour circumventing the cut on the real
negative axis can be represented as

\bea
&&\int\limits_{C_{\mbox{\DamirFont\symbol{'002}}}}d\omega{\cal
F}_2(\omega)
={\sin(z\pi)\over 2\pi^2(z-1)}\times\nonumber\\
&&\times
\int\limits_{m^2}^\infty {du\over u}(u-m^2)^{1-z}\bigl[
FI_F(r\sqrt{u})K_F(r\sqrt{u})+ 
(1-F)I_{1-F}(r\sqrt{u})K_{1-F}(r\sqrt{u})\bigr]
\label{C.6}
\eea
and
\bea
&&
\int\limits_{C_{\mbox{\DamirFont\symbol{'002}}}}d\omega{\cal
F}_3(\omega) =
{1\over2(2\pi)^2}\int\limits_{m^2}^\infty du(u-m^2)^{-{1\over2}-z}\times\nonumber\\
&&\times \biggl( A\mu^{1-2F}u^F\bigl\{
e^{i(F-z)\pi}\bigl[\tRp-\tRpm\bigr]- 
e^{-i(F-z)\pi}\bigl[ \tRmp-\tRm\bigr]\bigr\}I_{-F}(r\sqrt{u})K_F(r\sqrt{u})-\nonumber\\
&&-A\mu^{1-2F}u^{-1+F} \bigl\{ e^{i(F-z)\pi}
\bigl[(m-i\sqrt{u-m^2})^2\tRp- 
(m+i\sqrt{u-m^2})^2\tRpm\bigr]-\nonumber \\
&&-e^{-i(F-z)\pi}\bigl[
(m-i\sqrt{u-m^2})^2\tRmp- (m+i\sqrt{u-m^2})^2\tRm\bigr]\bigr\}
I_{1-F}(r\sqrt{u})K_{1-F}(r\sqrt{u})+\nonumber \\
&&+\bigl\{ e^{i(F-z)}\bigl[
(m+i\sqrt{u-m^2})\tRp-(m-i\sqrt{u-m^2})\tRpm\bigr] -\nonumber\\
&&-e^{-i(F-z)\pi}
\bigl[(m+i\sqrt{u-m^2})\tRmp-
(m-i\sqrt{u-m^2})\tRm\bigr] \bigr\} I_F(r\sqrt{u})K_F(r\sqrt{u})+\nonumber\\
&&+
\bigl\{e^{-i(F+z)\pi}\bigl[
(m+i\sqrt{u-m^2})\tRp-(m-i\sqrt{u-m^2})\tRpm\bigr\}-\nonumber\\
&&-e^{i(F+z)\pi} \bigl[(m+i\sqrt{u-m^2})\tRmp-
(m-i\sqrt{u-m^2})\tRm\bigr]\bigr\}
I_{-F}(r\sqrt{u})K_F(r\sqrt{u})-\nonumber \\
&&-\bigl\{e^{-i(F+z)\pi} \bigl[(m-i\sqrt{u-m^2})\tRp-
(m+i\sqrt{u-m^2})\tRpm\bigr]-\nonumber\\
&&-e^{i(F+z)\pi}\bigl[(m-i\sqrt{u-m^2})\tRmp-
(m+i\sqrt{u-m^2})\tRm\bigr]\bigr\}
I_{1-F}(r\sqrt{u})K_{1-F}(r\sqrt{u})-\nonumber\\
&&- \bigl\{ e^{i(F-z)\pi}\bigl[
(m-i\sqrt{u-m^2})\tRp-(m+i\sqrt{u-m^2})\tRpm\bigr]-\nonumber \\
&&- e^{-i(F-z)\pi}\bigl[(m-i\sqrt{u-m^2})\tRmp
-(m+i\sqrt{u-m^2})\tRm\bigr]\bigr\}
I_{-1+F}(r\sqrt{u})K_{1-F}(r\sqrt{u})+\nonumber \\
&&+ A^{-1}\mu^{2F-1}u^{-F}\bigl\{ e^{-i(F+z)\pi}
\bigl[(m+i\sqrt{u-m^2})^2\tRp- (m-i\sqrt{u-m^2})^2
\tRpm\bigr]-\nonumber \\
&&-e^{i(F+z)\pi}\bigl[ (m+i\sqrt{u-m^2})^2\tRmp-
(m-i\sqrt{u-m^2})^2\tRm\bigr]\bigr\}
I_F(r\sqrt{u})K_F(r\sqrt{u})-\nonumber \\
&&- A^{-1}\mu^{2F-1}u^{1-F}\bigl\{ e^{-i(F+z)\pi}
\bigl[\tRp-\tRpm\bigr] -e^{i(F+z)\pi}\bigl[\tRmp-\tRm\bigr]\bigr\}
I_{-1+F}(r\sqrt{u})K_{1-F}(r\sqrt{u})\biggr),
\label{C.7}
\eea
where
\bea
&&\tilde{R}_{(\pm)}^{(+)}=\bigg[ A\mu^{1-2F}u^{-1+F}e^{iF\pi}(m\mp
i\sqrt{u-m^2})+\nonumber\\
&&+2\cos(F\pi)+A^{-1}\mu^{2F-1}u^{-F} e^{-iF\pi}(m\pm
i\sqrt{u-m^2})\bigg]^{-1}
\label{C.8}
\eea
and
\bea
&&\tilde{R}_{(\pm)}^{(-)}=\bigl[ A\mu^{1-2F}u^{-1+F}e^{-iF\pi} (m\mp
i\sqrt{u-m^2})+\nonumber\\
&&+2\cos(F\pi)+ A^{-1}\mu^{2F-1}u^{-F}e^{iF\pi} (m\pm
i\sqrt{u-m^2})\bigr]^{-1}.
\label{C.9}
\eea
Adding the integral along the contour enclosing the pole of ${\cal
F}_2(\omega)$

\be
\int\limits_{C_{\mbox{\DamirFont\symbol{'011}}}}d\omega{\cal
F}_2(\omega) =
2\pi i\Res_{\omega=0}{\cal F}_2(\omega)
\label{C.10}
\ee
to Eq.(C.6), we arrive at Eq.(4.21) where variable $w=r\sqrt{u}$ is
introduced.

As to expression (C.7), it can be simplified
\bea
&& \int\limits_{C_{\mbox{\DamirFont\symbol{'002}}}}d\omega{\cal
F}_3(\omega) =
{\sin(z\pi)\over2\pi^2} \int\limits_{m^2}^\infty du(u-m^2)^{-z}
\bigl[I_F(r\sqrt{u})K_F(r\sqrt{u})+I_{1-F}(r\sqrt{u})K_{1-F}(r\sqrt{u})
\bigr]+\nonumber \\
&&+ {\sin(F\pi)\over4\pi^3}
\int\limits_{m^2}^\infty du(u-m^2)^{-{1\over2}-z}
\biggl(e^{-iz\pi}\bigl\{ \bigl[ A\mu^{1-2F}u^{F}e^{iF\pi}+
(m+i\sqrt{u-m^2})e^{-iF\pi}\bigr]\tRp-\nonumber \\
&&-\bigl[ A\mu^{1-2F}u^Fe^{iF\pi}+
(m-i\sqrt{u-m^2})e^{-iF\pi}\bigr]\tRpm\bigr\} -e^{iz\pi}\bigl\{
\bigl[A\mu^{1-2F}u^Fe^{-iF\pi}+\nonumber \\
&&+(m+i\sqrt{u-m^2}) e^{iF\pi}\bigr]\tRmp-\bigl[
A\mu^{1-2F}u^Fe^{-iF\pi}+ (m-i\sqrt{u-m^2})e^{iF\pi}\bigr]\tRm\bigr\}
\biggr)K_F^2(r\sqrt{u})-\nonumber \\
&&-{\sin(F\pi)\over4\pi^3} \int\limits_{m^2}^\infty
du(u-m^2)^{-{1\over2}-z}\biggl( e^{-iz\pi}\bigl\{\bigl[
A^{-1}\mu^{2F-1}u^{1-F}e^{-iF\pi}+
(m-i\sqrt{u-m^2})e^{iF\pi}\bigr]\tRp-\nonumber \\
&&-\bigl[ A^{-1}\mu^{2F-1}u^{1-F}e^{-iF\pi}+ (m+i\sqrt{u-m^2})e^{iF\pi}
\bigr]\tRpm\bigr\} -\nonumber\\
&&-e^{iz\pi}\bigl\{
\bigl[A^{-1}\mu^{2F-1}u^{1-F}e^{iF\pi}
+ (m-i\sqrt{u-m^2})e^{-iF\pi}\bigr]\tRmp-\nonumber\\
&&-
\bigl[A^{-1}\mu^{2F-1}u^{1-F}e^{iF\pi}+
(m+i\sqrt{u-m^2})e^{-iF\pi}\bigr]\tRm\bigr\}
\biggr)K_{1-F}^2(r(\sqrt{u}).
\label{C.11}
\eea
Simplifying further the last expression, we arrive at the terms in relation
(4.22) that are represented as integrals over variable
$w=r\sqrt{u}$.

It remains to consider the integral along the contour enclosing the
pole of ${\cal F}_3(\omega)$
\be
\int\limits_{C_{\mbox{\DamirFont\symbol{'011}}}}d\omega{\cal
F}_3(\omega) =
2\pi i\Res_{\omega=-\kappa^2} {\cal F}_3(\omega).
\label{C.12}
\ee
Choosing the branch for fractional exponents according to the
prescription
\be
(-\kappa^2)^\rho=\kappa^{2\rho}e^{i\rho\pi}, \qquad 0<\rho<1,
\label{C.13}
\ee
we get
\bea
&&\int\limits_{C_{\mbox{\DamirFont\symbol{'011}}}}d\omega{\cal
F}_3(\omega) =
\pm {1\over4\pi}\big|E_\BS\big|^{-1-2z}\bigl[
A\mu^{1-2F}\kappa^{2F}e^{iF\pi}I_{-F}(\kappa r)K_F(\kappa r)-\nonumber
\\
&&-A\mu^{1-2F}\kappa^{-2(1-F)}e^{iF\pi}(m\mp|E_\BS|)^2 I_{1-F}(\kappa
r)K_{1-F}(\kappa r)+e^{iF\pi}(m\pm|E_\BS|)I_F(\kappa r)K_F(\kappa
r)+\nonumber \\
&&+e^{-iF\pi} (m\pm|E_\BS|)I_{-F}(\kappa r)K_F(\kappa
r)-e^{-iF\pi}(m\mp|E_\BS|)I_{1-F}(\kappa r)K_{1-F}(\kappa r)-\nonumber
\\
&&-e^{iF\pi}(m\mp|E_\BS|)I_{-1+F}(\kappa r)K_{1-F}(\kappa
r)+A^{-1}\mu^{2F-1}\kappa^{-2F}e^{-iF\pi}(m\pm|E_\BS|)^2I_F(\kappa
r)K_F(\kappa r)-\nonumber \\
&&-A^{-1}\mu^{2F-1}\kappa^{2(1-F)}e^{-iF\pi}I_{-1+F}(\kappa
r)K_{1-F}(\kappa r)\bigr]
\Res_{\omega=-\kappa^2}\tilde{L}_{(\pm)}=\nonumber \\
&&=\mp{i\sin^2(F\pi)\over\pi^2} |E_\BS|^{-1-2z}\bigl[
(m\pm|E_\BS|)K_F^2(\kappa r)+(m\mp|E_\BS|)K_{1-F}^2(\kappa
r)\bigr]\Res_{\omega=-\kappa^2}\tilde{L}_{(\pm)},
\nonumber \\
&&\qquad \qquad \qquad E_\BS\gtrless 0.
\label{C.14}
\eea
Taking into account relation
\be
\Res_{\omega=-\kappa^2}\tilde{L}_{(\pm)}={1\over i\sin(F\pi)}\,
{|E_\BS|\kappa^2\over |E_\BS|(2F-1)\pm m}, \qquad E_\BS\gtrless 0,
\label{C.15}
\ee
we arrive at the last term in relation (4.22). Naturally, the result
will be the same, if the branch for fractional exponents is chosen
alternatively as
\be
(-\kappa^2)^\rho=\kappa^{2\rho}e^{-i\rho\pi}, \qquad 0<\rho<1.
\label{C.16}
\ee

%\cl
\section*{Appendix D}
\def\theequation{D.\arabic{equation}}
\setcounter{equation}{0}

Expression (4.30) is recast into the form
\be
[\zeta_\x(z|M)]_\IRREG=
\int\limits_{C_{\mbox{\DamirFont\symbol{'001}}}}d\omega{\cal
F}_4(\omega),
\label{D.1}
\ee
where
\bea
&&{\cal F}_4(\omega)= {1\over i2(2\pi)^2} (\omega+M^2)^{-z}\bigl\{
A\bigl({\omega\over\mu^2}\bigr)^{F-{1\over2}}
[\Lp-\Lm][I_{-F}(r\sqrt{-\omega})K_F(r\sqrt{-\omega})+\nonumber \\ 
&&+I_{1-F}(r\sqrt{-\omega})K_{1-F}(r\sqrt{-\omega})] +
\omega^F[\Lp+\Lm]
(-\omega)^{-F}I_F(r\sqrt{-\omega})K_F(r\sqrt{-\omega})+ 
\omega^{-F}[\Lp+\Lm]\times\nonumber \\
&&\times(-\omega)^F I_{-F}(r\sqrt{-\omega})K_F(r\sqrt{-\omega})
-\omega^{1-F}[\Lp+\Lm](-\omega)^{-1+F}
 I_{1-F}(r\sqrt{-\omega})K_{1-F}(r\sqrt{-\omega})-\nonumber \\ 
&&-\omega^{-1+F}[\Lp+\Lm](-\omega)^{1-F}
 I_{-1+F}(r\sqrt{-\omega})K_{1-F}(r\sqrt{-\omega})+\nonumber \\ 
&&+A^{-1}\bigl({\omega\over\mu^2}\bigr)^{{1\over2}-F}[\Lp-\Lm]
[I_F(r\sqrt{-\omega})K_F(r\sqrt{-\omega})+I_{-1+F}(r\sqrt{-\omega})
K_{1-F}(r\sqrt{-\omega})]\bigr\}.
\label{D.2}
\eea
Deforming the contour of integration in Eq.(D.1), we arrive at the
same relation as Eq.(B.4), but with contour
$C_{\mbox{\DamirFont\symbol{'002}}}$
circumventing the cut from $-M^2$ to $-\infty$ (see Fig.4). At
${1\over2}<\Re z<1$ the integrals along semicircles
$C_{\mbox{\DamirFont\symbol{'005}}}$ and
$C_{\mbox{\DamirFont\symbol{'004}}}$ of infinite radii 
vanish, so there remains the integral:
\bea
&&
\int\limits_{C_{\mbox{\DamirFont\symbol{'002}}}}d\omega{\cal
F}_4(\omega) =
{1\over 2(2\pi)^2}\int\limits_{M^2}^\infty du(u-M^2)^{-z} \biggl(
A\bigl({u\over \mu^2}\bigr)^{F-{1\over2}} \bigl\{
e^{i(F-z)\pi}[\Rp-\Rpm]-\nonumber \\
&&-e^{-i(F-z)\pi}[\Rmp-\Rm]\bigr\} \bigl[
I_{-F}(r\sqrt{u})K_F(r\sqrt{u})+I_{1-F}(r\sqrt{u})
K_{1-F}(r\sqrt{u})\bigr]-\nonumber \\
&&-\bigl\{ e^{i(F-{1\over2}-z)\pi}[\Rp+\Rpm]+
e^{-i(F-{1\over2}-z)\pi}[\Rmp+\Rm]\bigr\}
[I_F(r\sqrt{u})K_F(r\sqrt{u})+\nonumber \\
&&+I_{-1+F}(r\sqrt{u})K_{1-F}(r\sqrt{u})]+\bigl\{
e^{-i(F-{1\over2}+z)\pi}[\Rp+\Rpm]+e^{i(F-{1\over2}+z)\pi}[\Rmp+\Rm]
\bigr\}\times\nonumber \\
&&\times [I_{-F}(r\sqrt{u})K_F(r\sqrt{u})+
I_{1-F}(r\sqrt{u})K_{1-F}(r\sqrt{u})]-
A^{-1}\bigl({u\over\mu^2}\bigr)^{{1\over2}-F}\bigl\{ e^{-i(F+z)\pi}
[\Rp-\Rpm]-\nonumber \\
&&-e^{i(F+z)\pi}[\Rmp-\Rm]\bigr\}
[I_F(r\sqrt{u})K_F(r\sqrt{u})+I_{-1+F}(r\sqrt{u})K_{1-F}(r\sqrt{u})]\biggr),
\label{D.3}
\eea
where $R_{(\pm)}^{(+)}$ and $R_{(\pm)}^{(-)}$ are given by Eqs.(B.6)
and (B.7). Expression (D.3) can be reduced to the form

\bea
&& \int\limits_{C_{\mbox{\DamirFont\symbol{'002}}}}d\omega{\cal
F}_4(\omega) =
{\sin(z\pi)\over2\pi^2} \int\limits_{M^2}^\infty du(u-M^2)^{-z}
[I_F(r\sqrt{u})K_F(r\sqrt{u})+
I_{1-F}(r\sqrt{u})K_{1-F}(r\sqrt{u})]+\nonumber \\
&&+{\sin(F\pi)\over4\pi^3} \int\limits_{M^2}^\infty du(u-M^2)^{-z}
\bigl\{ A\bigl({u\over\mu^2}\bigr)^{F-{1\over2}}\bigl[
e^{i(F-z)\pi}(\Rp-\Rpm) -e^{-i(F-z)\pi} (\Rmp-\Rm)\bigr]+\nonumber \\
&& + e^{-i(F-{1\over2}+z)\pi} (\Rp+\Rpm)+ e^{i(F-{1\over2}+z)\pi}
(\Rmp+\Rm)\bigr\} K_F^2(r\sqrt{u})-\nonumber \\
&&-{\sin(F\pi)\over4\pi^3} \int\limits_{M^2}^\infty du (u-M^2)^{-z}
\bigl\{e^{i(F-{1\over2}-z)\pi} (\Rp+\Rpm)+ e^{-i(F-{1\over2}-z)\pi}
(\Rmp+\Rm)+\nonumber \\
&&+ A^{-1}\bigl({u\over\mu^2}\bigr)^{{1\over2}-F}\bigl[
e^{-i(F+z)\pi}(\Rp-\Rpm)- e^{i(F+z)\pi}(\Rmp-\Rm)\bigr]\bigr\}
K_{1-F}^2(r\sqrt{u}).
\label{D.4}
\eea
After further simplifications we arrive at Eq.(4.31) where variable
$w=r\sqrt{u}$ is introduced.

\newpage

\begin{figure}[ht]
\centerline{\epsfig{figure=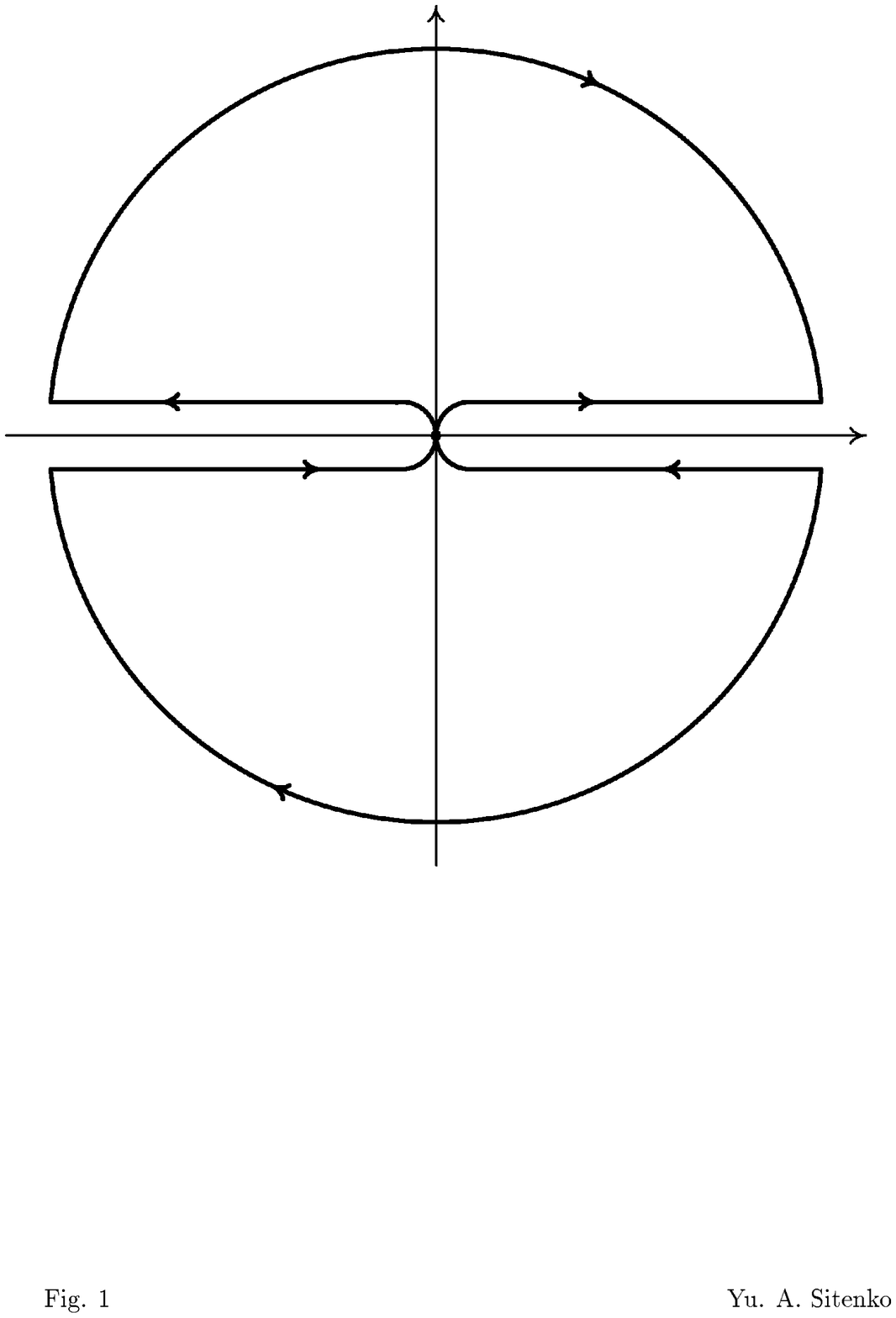,width=\textwidth}}
\end{figure}

\newpage

\begin{figure}[ht]
\centerline{\epsfig{figure=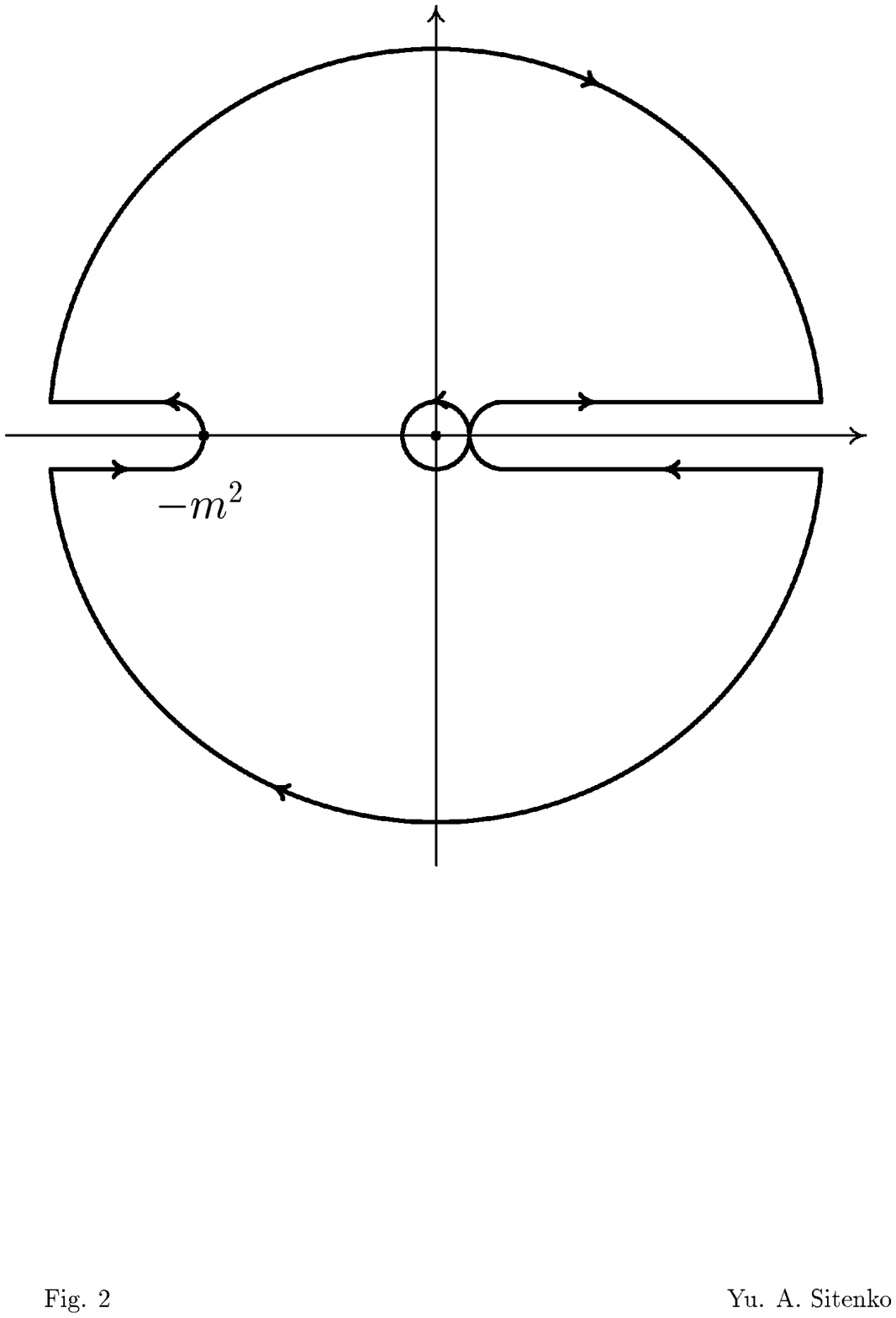,width=\textwidth}}
\end{figure}

\newpage

\begin{figure}[ht]
\centerline{\epsfig{figure=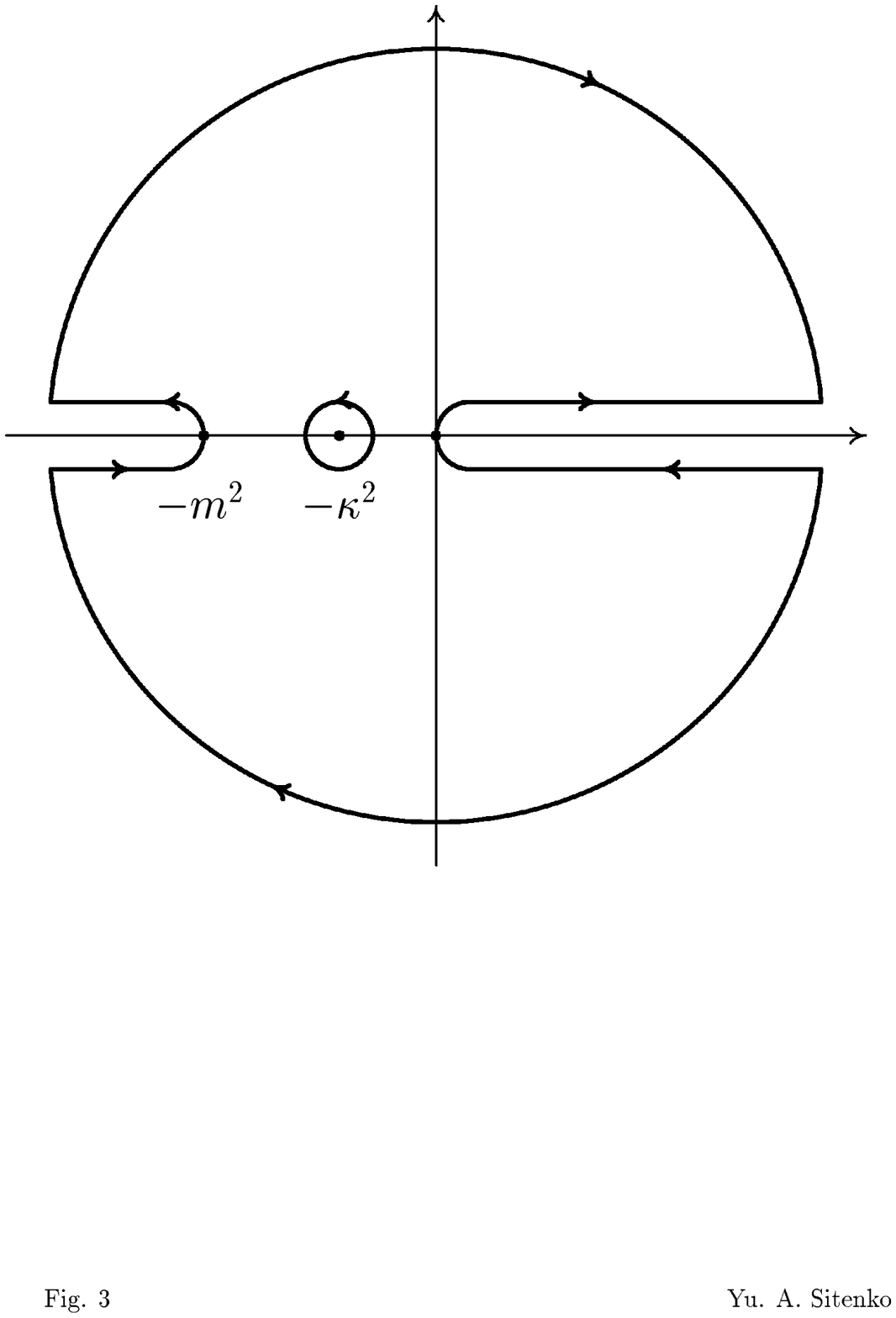,width=\textwidth}}
\end{figure}

\newpage

\begin{figure}[ht]
\centerline{\epsfig{figure=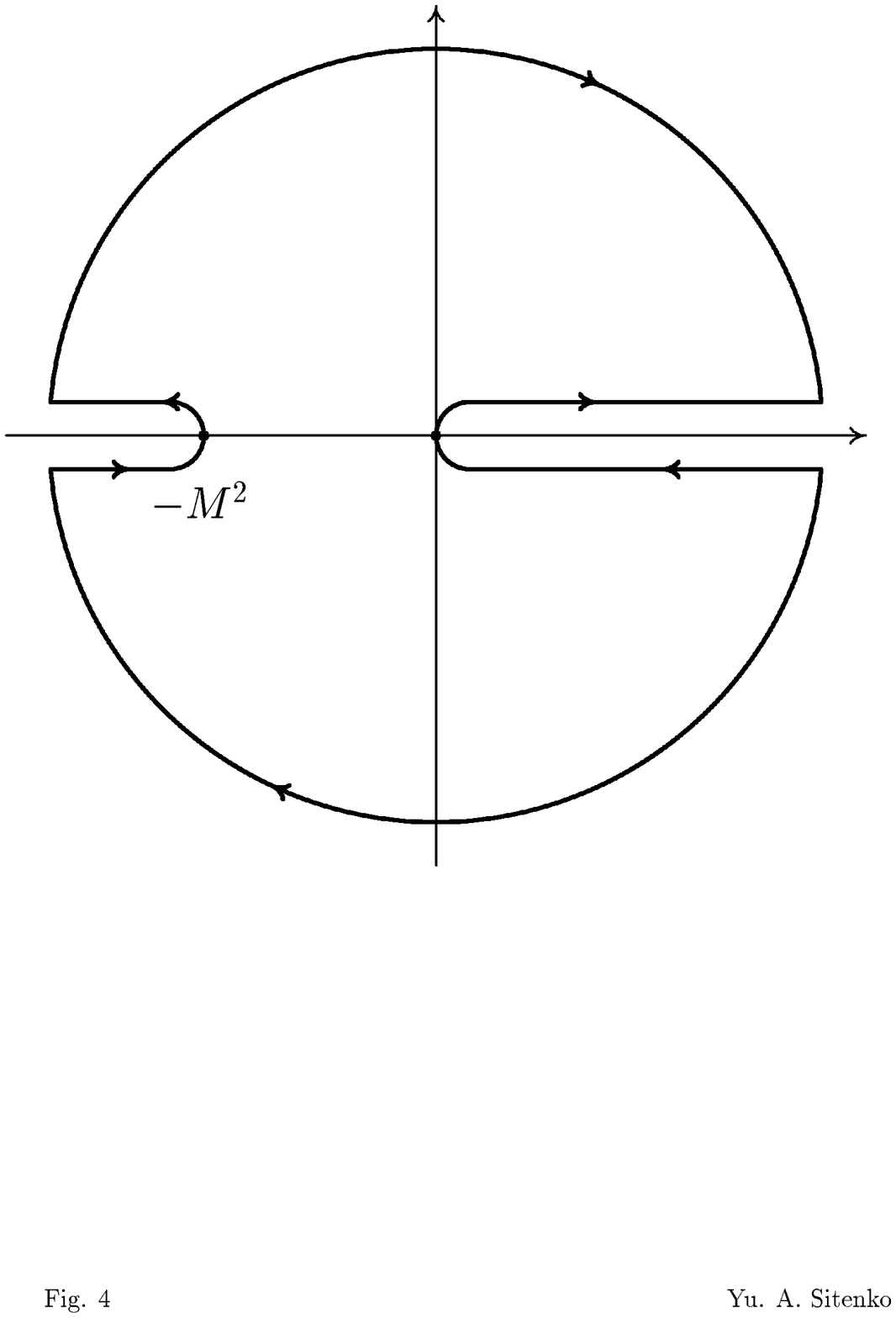,width=\textwidth}}
\end{figure}

\end{document}